\documentclass[aos]{imsart}

\RequirePackage[OT1]{fontenc}
\RequirePackage{amsthm,amsmath}
\RequirePackage[colorlinks,citecolor=blue,urlcolor=blue]{hyperref}

\usepackage{graphicx,psfrag,epsf,subcaption}
\usepackage{algorithm}
\usepackage{algpseudocode}%
\usepackage{enumerate}
\usepackage{enumitem}
\usepackage[sort&compress]{natbib}
\usepackage{amssymb}
\usepackage{bm}
\usepackage{url}
\usepackage{float}
\usepackage{siunitx}
\usepackage{bbm}
\usepackage{booktabs}
\usepackage{array}
\usepackage{comment}
\usepackage{anyfontsize}

\usepackage[toc,page]{appendix}
\usepackage[font=small,labelfont=bf]{caption}
\RequirePackage[normalem]{ulem} 

\graphicspath{{plots/}}

\startlocaldefs

\let\hat\widehat

\ifx\BlackBox\undefined
\newcommand{\BlackBox}{\rule{1.5ex}{1.5ex}}  
\fi

\ifx\QED\undefined
\def\QED{~\rule[-1pt]{5pt}{5pt}\par\medskip}
\fi

\ifx\proof\undefined
\newenvironment{proof}{\par\noindent{\bf Proof\ }}{\hfill\BlackBox\\[2mm]}
\fi

\ifx\theorem\undefined
\newtheorem{theorem}{Theorem}
\fi
\ifx\example\undefined
\newtheorem{example}{Example}
\fi
\ifx\property\undefined
\newtheorem{property}{Property}
\fi
\ifx\lemma\undefined
\newtheorem{lemma}[theorem]{Lemma}
\fi
\ifx\proposition\undefined
\newtheorem{proposition}[theorem]{Proposition}
\fi
\ifx\remark\undefined
\newtheorem{remark}{Remark}
\fi
\ifx\corollary\undefined
\newtheorem{corollary}[theorem]{Corollary}
\fi
\ifx\definition\undefined
\newtheorem{definition}[theorem]{Definition}
\fi
\ifx\conjecture\undefined
\newtheorem{conjecture}[theorem]{Conjecture}
\fi
\ifx\fact\undefined
\newtheorem{fact}[theorem]{Fact}
\fi
\ifx\claim\undefined
\newtheorem{claim}[theorem]{Claim}
\fi
\ifx\assumption\undefined

\fi
\numberwithin{equation}{section}
\numberwithin{theorem}{section}
\usepackage{multirow}

\newcommand{\cF}{\mathcal{F}}

\newcommand{\E}{\mathbb{E}}
\newcommand{\bx}{\mathbf{x}}

\renewcommand{\Pr}{\mathbb{P}}

\newcommand{\cX}{\mathcal{X}}
\newcommand{\cY}{\mathcal{Y}}

\newcommand{\cH}{\mathcal{H}}

\newcommand{\var}{\mbox{Var}}

\newcommand{\cP}{\mathcal{P}}

\newcommand{\bbR}{\mathbb{R}}

\newcommand{\bD}{\mathbb{D}}

\newcommand{\cD}{\mathcal{D}}

\newcommand{\cN}{\mathcal{N}}

\def\T{^{ \mathrm{\scriptscriptstyle T} }}

\newcommand{\argmin}{\mathop{\mathrm{argmin}}}

\setlist[itemize]{leftmargin=.3in}

\newtheorem{assumption}{Assumption}

\setlist{leftmargin=15mm}
\endlocaldefs

\begin{document}
\begin{frontmatter}
\title{Bootstrap Nonparametric Inference under Data Integration}
\runtitle{Bootstrap Nonparametric Inference under Data Integration}

\begin{aug}
\author[A]{\fnms{Zuofeng}~\snm{Shang}\ead[label=e1]{zshang@njit.edu}},
\author[B]{\fnms{Peijun}~\snm{Sang}\ead[label=e2]{peijun.sang@uwaterloo.ca}}
\and
\author[A]{\fnms{Chong}~\snm{Jin}\ead[label=e3]{chong.jin@njit.edu}}
\address[A]{Department of Mathematical Sciences, New Jersey Institute of Technology, 
\printead{e1}}

\address[B]{Department of Statistics and Actuarial Science, University of Waterloo,
\printead{e2}}

\address[A]{Department of Mathematical Sciences, New Jersey Institute of Technology, 
\printead{e3}}

\end{aug}

\begin{abstract}
We propose multiplier bootstrap procedures for nonparametric inference and uncertainty quantification of the target mean function, based on a novel framework of integrating target and source data. 
We begin with the relatively easier covariate shift scenario with equal target and source mean functions
and propose estimation and inferential procedures through a straightforward combination of all target and
source datasets.
We next consider the more general and flexible distribution shift scenario with arbitrary target and source mean functions,
and propose a two-step inferential procedure.
First, we estimate the target-to-source differences based on separate portions of the target and source data. Second, the remaining source data are adjusted by these differences and combined with the remaining target data to perform the multiplier bootstrap procedure.
Our method enables local and global inference on the target mean function without using asymptotic distributions. To justify our approach, we derive an optimal convergence rate for the nonparametric estimator and establish bootstrap consistency to estimate the asymptotic distribution of the nonparametric estimator. The proof of global bootstrap consistency involves a central limit theorem for quadratic forms with dependent variables under a conditional probability measure.
Our method applies to arbitrary source and target datasets, provided that the data sizes meet a specific quantitative relationship.
Simulation studies and real data analysis are provided to examine the performance of our approach.
\end{abstract}

\begin{keyword}[class=MSC]
\kwd[Primary ]{62G20}
\kwd[; secondary ]{62G05}
\end{keyword}

\begin{keyword}
\kwd{nonparametric inference}
\kwd{uncertainty quantification}
\kwd{multiplier bootstrap}
\kwd{bootstrap consistency}
\kwd{data integration}
\end{keyword}

\end{frontmatter}

\section{Introduction}

In scientific applications, data are often limited in a certain target task.
Examples include healthcare or clinical study, retail and e-commerce, financial services, etc.,
in which the target data are often limited due to safety, privacy, or regulatory concerns.
Traditional statistical methods often perform unsatisfactorily due to data insufficiency. 
The state-of-the-art data integration architecture (see \citealt{halevy2001answering, hull1997managing, ullman1997information, maurizio2002amc}) aims at reusing and adapting the precollected data from related source tasks to enhance performance on the target tasks and therefore has become a fundamental technique in numerous fields, offering the potential to improve statistical performance while reducing the need for large amounts of target data. 
This article aims to explore nonparametric inference
and uncertainty quantification under the framework of data integration, which, to the best of our knowledge, have remained unresolved challenges in the literature.

Data integration requires the entire dataset to consist of one target dataset and multiple source datasets. Estimation and inferential procedures shall be proposed for the nonparametric mean function of the target data by efficiently integrating the target and source datasets.
Our proposed framework allows the target and
source data sets to be arbitrarily dissimilar, which
is different and more flexible than two relevant frameworks: distributed learning and transfer learning. In distributed learning, data from different units are typically assumed to have equal distributions (\citealt{zhang2013jmlr, shang2017jmlr, shang2019jmlr, banerjee2019aos, shang:ejs:2020, szabo2020aos}). Transfer learning is somewhat more flexible, allowing for different target and source distributions, though they must be sufficiently similar to ensure data transferability \citep{li2021jrssb,li2023jasa,li2024jasa,ye2023jasa,cai2021aos,cai2024arxiv,wang2023minimaxoptimaltransferlearning}. 
The proposed framework for data integration, on the contrary, aims to integrate arbitrary target and source datasets, even when they are not necessarily similar, thereby overcoming the similarity constraint.
Despite this clear advantage, performing nonparametric inference for the target mean function remains a challenge.  
Specifically, classic nonparametric inference often requires deriving
the asymptotic distribution of the nonparametric estimator, which is hard to achieve under integrated data. Moreover, the asymptotic distribution, even if it is feasible, often involves nuisance parameters for which estimation could be complicated. 
We plan to address these challenges through a two-stage multiplier bootstrap method and establish its statistical validity. 

We briefly summarize the two-stage method as follows.
The first stage is to divide the target and source datasets into equal-sized portions and use the initial portions to estimate the source-to-target mean differences. The second stage is to use these estimated differences to calibrate the remaining portions before they are integrated to perform multiplier bootstrap inference. Our method can perform local and global inferences on the target mean function
without resorting to the asymptotic distribution of the nonparametric estimator.
In particular, we incorporate multiplier weights into the second stage of the method to generate multiple bootstrap estimators of the target mean function. We will construct pointwise confidence intervals and global confidence regions for the target mean function directly based on these bootstrap estimators, which avoids the need of an asymptotic distribution of the estimator.

A novel aspect of this article is to establish the statistical guarantee of our method when integrating arbitrarily dissimilar target and source datasets, as long as the sizes of these datasets satisfy a specific quantitative relationship. This relationship is crucial to ensure the effectiveness of our method. Intuitively, if the target data size is considerably smaller than that of the source data, the source data may exert excessive influence, skewing the results. Conversely, if the target data size is substantially larger, integrating the source data is unlikely to provide significant improvements over using only the target data.
Therefore, the target and source data sizes must be balanced to ensure that the integrated data yield asymptotically valid and more accurate results. We will precisely quantify this relationship when the mean functions belong to various concrete function spaces, providing theoretical insights into how to balance target and source datasets to make our approach work.

The contributions of this paper are summarized in more detail as follows.
First, we propose a nonparametric estimator and local/global multiplier bootstrap inferential procedures
for the target mean function in both covariate shit and distribution shift scenarios.
In the former, the target and source mean functions are identical, which is a simpler scenario to begin with. 
We will estimate the target mean through a straightforward combination of all target and source datasets. 
In the latter case, the target and source mean functions are general and arbitrary. We will initially estimate the source-to-target differences through data splitting, which are used to calibrate the remaining source data.
The calibrated source data and remaining target data are then integrated to perform multiplier bootstrap estimation of the target mean function. We further construct pointwise confidence intervals and a global confidence region for the target mean function based on
these bootstrap estimators.
Second, we derive an optimal rate of convergence for the proposed estimator,
as well as bootstrap consistency in estimating the limiting distribution of the estimator.
Based on the bootstrap consistency, we show that the proposed local/global inferential procedures
achieve an asymptotically desirable coverage probability.
Furthermore, we establish quantified relationships between the sizes of target and source data
such that our theoretical results hold under arbitrary target and source datasets without requiring their similarity. This is a great breakthrough compared to the existing data integration frameworks such as distributed learning and transfer learning. 
Third, through simulation studies, we examine the performance of our method versus various ratios of the source data size to the target data size. The results demonstrate that our method may outperform the ones using only target data, and may underperform when the ratio becomes significantly large; these are consistent with our theoretical finding.
Moreover, our method can advance real-world applications such as nonparametric inference and uncertainty quantification in Alzheimer's disease data. By leveraging the integrated data, we aim to improve statistical accuracy compared with current methods using
only target data, thus enhancing the inferential results in the Alzheimer’s disease study.

The rest of this article is organized as follows.
In Section \ref{sec:methods}, we propose estimation and inference methods for the nonparametric mean function using integrated data,
in both covariate shift and distribution shift scenarios.
In Section \ref{sec:assumptions}, we introduce a reproducing kernel Hilbert space (RKHS) framework as well as technical assumptions,
which are useful for asymptotic analysis of the proposed methods.
In Section \ref{sec:asymptotic}, we provide the asymptotic analysis for the proposed methods such as the optimal convergence rate for the nonparametric estimator and bootstrap consistency
for estimating the asymptotic distribution of the nonparametric estimator.
Section \ref{sec:sim} provides simulation studies to examine the finite-sample performance of our method. In Section \ref{sec:realdata} we apply our methods to analyze Alzheimer’s disease data. We make some concluding remarks in Section \ref{sec:conclusions}, and the proofs of the main theorems in Section \ref{sec:asymptotic} are collected in Section \ref{sec:proofs}.

\section{Estimation and inference for mean function under data integration}\label{sec:methods}

Suppose that $(Y^{(0)},X^{(0)})\in\cY\times\cX$ is a target variable generated from an unknown target distribution $P^{(0)}$,
where $Y^{(0)}\in\cY$ is a response variable, $X^{(0)}\in\cX$ is a predictor variable, and $\cY\subset\bbR$ and
$\cX\subset\bbR^d$ are real subsets. 
Our aim is to estimate the conditional mean function 
\[
\textrm{$f^{(0)}_*(x)=\E[Y^{(0)} \mid X^{(0)}=x]$,\,\,\,\, for $x\in\cX$.}
\]
A common practice is to use the target data $(Y_1^{(0)},X_1^{(0)}),\ldots,(Y_{n_0}^{(0)},X_{n_0}^{(0)})$, which are $n_0$
\textit{iid} copies of $(Y^{(0)},X^{(0)})$, to achieve this purpose.
There is a rich literature in the field, so we only list several representative textbooks, e.g., smoothing spline regression \citep{w90, gu2013, wang2011},
local polynomial regression \citep{fan1996local} and sieve estimation \citep{CHEN20075549}.

Additionally, suppose that there are source variables $(Y^{(1)},X^{(1)}),\ldots,(Y^{(M)},X^{(M)})\in\cY\times\cX$ 
generated from unknown source distributions
$P^{(1)},\ldots,P^{(M)}$, respectively. 
Suppose that, for $1\le m\le M$, we observe source data
$(Y_1^{(m)},X_1^{(m)}),\ldots,(Y_{n_m}^{(m)},X_{n_m}^{(m)})$, which are $n_m$ \textit{iid} copies of $(Y^{(m)},X^{(m)})$. 
It is of primary interest to integrate these source data into the target data to further enhance the estimation and statistical inference of $f^{(0)}_*$. 
The major challenge of this problem is that the conditional mean function of the $m$-th source, 
$f^{(m)}_*(x)=\E[Y^{(m)}|X^{(m)}=x]$, is not necessarily identical to $f^{(0)}_*$. Hence, a direct combination of the source data and target data generally does not work. We will study this problem in two situations: covariate shift and distribution shift, as illustrated in the following sections.

For convenience, we assume that $f^{(0)}_*$ belongs to a reproducing kernel Hilbert space $\cH$ of functions on $\cX$ endowed with a reproducing kernel $R(\cdot,\cdot)$ and an inner product $\langle\cdot,\cdot\rangle_\cH$. Therefore, by the reproducing property of $R$, we have $f^{(0)}_*(x)=\langle f^{(0)}_*, R_x\rangle_\cH$ for any $x\in\cX$, where $R_x(\cdot)\equiv R(x,\cdot)$ is an element of $\cH$. 
For $m = 1, \ldots, M$, let $e^{(m)}=Y^{(m)}-f^{(m)}_*(X^{(m)})$ be the model error of the $m$-th source, which has mean zero conditional on $X^{(m)}$.

\subsection{Covariate shift}\label{sec:cs}
In covariate shift, the conditional distributions of $Y^{(m)}$ given $X^{(m)}$ are identical,
i.e., $P^{(m)}(y|x)=P^{(0)}(y|x)$ for all $1\le m\le M$, while the marginal distributions of $X^{(m)}$, $P^{(m)}(x)$'s, might be different. Hence, $f^{(m)}_*(x)$ is identical to $f^{(0)}_*(x)$.
This means that $Y^{(0)},\ldots,Y^{(M)}$ have the same ``center'' $f^{(0)}_*$ so that we can directly integrate source data into target data to estimate $f^{(0)}_*$, which is similar in spirit to the distributed learning framework studied in the literature \citep{zhang2013jmlr, shang2017jmlr, shang2019jmlr, shang:ejs:2020, szabo2020aos}.
Specifically, we propose the following weighted penalized least squares regression for estimating $f^{(0)}_*$ via aggregating target and source data sets:
\begin{equation}\label{pmle:cs}
\widehat{f}^{\textrm{cs},\omega}_{n\lambda}=\argmin_{f\in\cH}\ell^{\textrm{cs},\omega}_{n\lambda}(f)\equiv\argmin_{f\in\cH}\left\{\frac{1}{2n}\sum_{m=0}^M\sum_{i=1}^{n_m}\omega_i^{(m)}|Y_i^{(m)}-f(X_i^{(m)})|^2+\frac{\lambda}{2}\|f\|_\cH^2\right\},
\end{equation}
where $\lambda>0$ is the smoothing parameter, $n=\sum_{m=0}^M n_m$ is the size of aggregated data, $\|f\|_\cH=\sqrt{\langle f,f\rangle_\cH}$ is the norm of $f$, and $\omega=\{\omega_i^{(m)}: m=0,\ldots,M, i=1,\ldots,n_m\}$ are either deterministic or random weights independent of the data. 
The superscript ``cs'' represents the term ``covariate shift.''
Throughout this article we only consider two cases: (1) $\omega_i^{(m)}=1$ for all $m,i$, or (2) $\omega_i^{(m)}$'s are i.i.d. bounded positive-valued random variables
with mean one and variance one. For instance, to satisfy (2), we can generate the weights from probability density $p(w)$, where
$p(w)=3/4$, if $0\le w\le 1$; $=1/12$, if $1<w\le 4$; and zero otherwise.

\subsection{Distribution shift}
In distribution shift, the joint distributions of the data variables,
$P^{(0)}(y,x),\ldots,P^{(M)}(y,x)$, are not necessarily identical. This means that the 
conditional distributions $P^{(0)}(y|x),\ldots,P^{(M)}(y|x)$ are not necessarily identical as well, so aren't the conditional mean functions
$f^{(0)}_*,\ldots,f^{(M)}_*$. Throughout this article, we do not require these mean functions to be identical or similar;
literally they can be arbitrarily unrelated. This framework is more general than distributed learning 
\citep{zhang2013jmlr, shang2017jmlr, shang2019jmlr, banerjee2019aos, shang:ejs:2020, szabo2020aos} and transfer learning \citep{cai2024arxiv,wang2023minimaxoptimaltransferlearning};
recall that transfer learning requires the distance between $f^{(0)}_*$ and $f^{(1)}_*,\ldots,f^{(M)}_*$ to be small so that the source data can be effectively transferred to estimate $f^{(0)}_*$, which is not needed here.
To take the heterogeneity into consideration when leveraging the source data to estimate $f^{(0)}_*$, 
an additional step is entailed to calibrate the centers of the source data. 
For this purpose, 
assume that we have additional target data $(Y_{n_0+1}^{(0)},X_{n_0+1}^{(0)}),\ldots,(Y_{2n_0}^{(0)},X_{2n_0}^{(0)})$,
which are $n_0$ \textit{iid} copies of $(Y^{(0)},X^{(0)})$,
as well as additional source data $(Y_{n_m+1}^{(m)},X_{n_m+1}^{(m)}),\ldots,(Y_{2n_m}^{(m)},X_{2n_m}^{(m)})$, which are $n_m$ \textit{iid} copies of $(Y^{(m)},X^{(m)})$, for $1 \leq m \leq M$. Moreover, we assume that these additional data are independent of the original data. 

We estimate $f^{(0)}_*$ in two steps. In Step 1, we estimate the difference $\delta^{(m)}:=f^{(m)}_*-f^{(0)}_*$, for $m=1,\ldots,M$, based on the original source and target data. These differences will be used to calibrate the centers of the source data.
In Step 2, we estimate $f^{(0)}_*$ with some ``pseudo'' source data calibrated by the estimation of $\delta^{(m)}$'s.

\begin{itemize}
    \item[\underline{Step 1.}] For $m=0,\ldots,M$, estimate $f^{(m)}_*$ through the following kernel ridge regression:
    \[
    \widehat{f}^{(m)}=\argmin_{f\in\cH}\ell^{(m)}(f)\equiv\argmin_{f\in\cH}\left\{\frac{1}{2n_m}\sum_{i=1}^{n_m}|Y_i^{(m)}-f(X_i^{(m)})|^2+\frac{\lambda}{2}\|f\|_\cH^2\right\},
    \]
    where $\lambda>0$ is the smoothing parameter. Calculate $\widehat{\delta}^{(m)}=\widehat{f}^{(m)}-\widehat{f}^{(0)}$ for $m=1,\ldots,M$ and set $\widehat{\delta}^{(0)}$=0.

    \item[\underline{Step 2.}] For $0\le m\le M$ and $n_m+1\le i\le 2n_m$, calculate $\widetilde{Y}_i^{(m)}=Y_i^{(m)}-\widehat{\delta}^{(m)}(X_i^{(m)})$. Estimate $f^{(0)}_*$ through the following weighted kernel ridge regression:
    \begin{equation}\label{pmle:ds}
    \widehat{f}^{\textrm{ds},\omega}_{n\lambda}=\argmin_{f\in\cH}\ell^{\textrm{ds},\omega}_{n\lambda}(f)\equiv\argmin_{f\in\cH}\left\{\frac{1}{2n}\sum_{m=0}^M\sum_{i=n_m+1}^{2n_m}\omega_i^{(m)}|\widetilde{Y}_i^{(m)}-f(X_i^{(m)})|^2+\frac{\lambda}{2}\|f\|_\cH^2\right\},    
    \end{equation}
    where $\lambda>0$ is the smoothing parameter, $n=\sum_{m=0}^M n_m$ and $\omega=\{\omega_i^{(m)}: m=0,\ldots,M, i=n_m+1,\ldots,2n_m\}$ are either deterministic or random positive weights independent of the data (similar to Section \ref{sec:cs}). The superscript ``ds'' represents the term ``distribution shift.''
\end{itemize}
Note that we have advocated the same $\lambda$ in both Steps 1 and 2, which is convenient for both methodological and theoretical development.

\subsection{Local and global bootstrap inference}\label{sec:inference:procedures}

We are interested in the following local and global inference problems. First, for any $x\in\cX$, how to construct a $100(1-\alpha)\%$ confidence interval for $f^{(0)}_*(x)$? Second, how to construct a $100(1-\alpha)\%$ confidence region for $f^{(0)}_*$? In this section, we propose bootstrap procedures to address both problems under covariate shift and distribution shift, respectively.

Let $\widehat{f}^{\textrm{cs}}_{n\lambda}$ and $\widehat{f}^{\textrm{ds}}_{n\lambda}$ be the estimated regression functions obtained by solving minimization problems
(\ref{pmle:cs}) and (\ref{pmle:ds}), respectively, with the weights therein all being ones.
Generate $B$ sets of bootstrap weights $\omega_1,\ldots,\omega_B$, with $\omega_b=\{\omega_{b,i}^{(m)}: m=0,\ldots,M,i=1,\ldots,n_m\}$ consisting of i.i.d.
bounded positive variables with mean one and variance one.
For $b=1,\ldots,B$, perform (\ref{pmle:cs}) and (\ref{pmle:ds}) to produce weighted kernel ridge estimates $\widehat{f}^{\textrm{cs},\omega_b}$ and $\widehat{f}^{\textrm{ds},\omega_b}$.

For any $x\in\cX$, we construct a $100(1-\alpha)\%$ confidence interval for $f^{(0)}_*(x)$ as follows:
\begin{itemize}
    \item[\underline{Step 1}.]  Let $p^{\textrm{cs}}_\alpha(x)$ and $q^{\textrm{cs}}_\alpha(x)$ be the lower $100(\alpha/2)\%$ and upper $100(1-\alpha/2)\%$ percentiles of 
    $\widehat{f}^{\textrm{cs},\omega_1}_{n\lambda}(x)-\widehat{f}^{\textrm{cs}}_{n\lambda}(x),\ldots,\widehat{f}^{\textrm{cs},\omega_B}_{n\lambda}(x)-\widehat{f}^{\textrm{cs}}_{n\lambda}(x)$, and let 
    $p^{\textrm{ds}}_\alpha(x)$ and $q^{\textrm{ds}}_\alpha(x)$ be the lower $100(\alpha/2)\%$ and upper $100(1-\alpha/2)\%$ percentiles of 
    $\widehat{f}^{\textrm{ds},\omega_1}_{n\lambda}(x)-\widehat{f}^{\textrm{ds}}_{n\lambda}(x),\ldots,\widehat{f}^{\textrm{ds},\omega_B}_{n\lambda}(x)-\widehat{f}^{\textrm{ds}}_{n\lambda}(x)$.
    \item[\underline{Step 2}.] Find $\textrm{CI}^{\textrm{cs}}_\alpha(x)=(\widehat{f}^{\textrm{cs}}_{n\lambda}(x)-q^{\textrm{cs}}_\alpha,\widehat{f}^{\textrm{cs}}_{n\lambda}(x)-p^{\textrm{cs}}_\alpha)$ and
    $\textrm{CI}^{\textrm{ds}}_\alpha(x)=(\widehat{f}^{\textrm{ds}}_{n\lambda}(x)-q^{\textrm{ds}}_\alpha,\widehat{f}^{\textrm{ds}}_{n\lambda}(x)-p^{\textrm{ds}}_\alpha)$.
\end{itemize}

For $f\in\cH$, define the empirical $L_2$-norm:
\[
\|f\|_{\textrm{ep}}^2=\frac{1}{n}\sum_{m=0}^M\sum_{i=1}^{n_m}f(X_i^{(m)})^2.
\]
We construct a $100(1-\alpha)\%$ confidence region for $f^{(0)}_*$ as follows:
\begin{itemize}
    \item[\underline{Step 1}.] Let $r^{\textrm{cs}}_\alpha$ be the upper $100(1-\alpha)\%$ percentile of 
$\|\widehat{f}^{\textrm{cs},\omega_1}_{n\lambda}-\widehat{f}^{\textrm{cs}}_{n\lambda}\|_{\textrm{ep}},\ldots,\|\widehat{f}^{\textrm{cs},\omega_B}_{n\lambda}-\widehat{f}^{\textrm{cs}}_{n\lambda}\|_{\textrm{ep}}$, and let $r^{\textrm{ds}}_\alpha$ be the upper $100(1-\alpha)\%$ percentile of $\|\widehat{f}^{\textrm{ds},\omega_1}_{n\lambda}-\widehat{f}^{\textrm{ds}}_{n\lambda}\|_{\textrm{ep}},\ldots,\|\widehat{f}^{\textrm{ds},\omega_B}_{n\lambda}-\widehat{f}^{\textrm{ds}}_{n\lambda}\|_{\textrm{ep}}$.
    \item[\underline{Step 2}.] Find $R^{\textrm{cs}}_\alpha=\{f\in\cH: \|f-\widehat{f}^{\textrm{cs}}_{n\lambda}\|_{\textrm{ep}}\le r^{\textrm{ds}}_\alpha\}$ and 
    $R^{\textrm{ds}}_\alpha=\{f\in\cH: \|f-\widehat{f}^{\textrm{ds}}_{n\lambda}\|_{\textrm{ep}}\le r^{\textrm{ds}}_\alpha\}$.
\end{itemize}

\section{Assumptions and an RKHS framework}\label{sec:assumptions}
In this section, we establish an RKHS framework under which our theoretical results will be obtained.
We first introduce several technical assumptions.
For $m=0,\ldots,M$ and $f,g\in\cH$, define quadratic forms
\[
\textrm{$V_m(f,g)=\E[f(X^{(m)})g(X^{(m)})]$ and $V(f,g)=\frac{1}{n}\sum_{m=0}^M n_m V_m(f,g)$,}
\]
where $n=\sum_{m=0}^M n_m$.
We can view $V_0$ as the quadratic form with respect to the target measure $P^{(0)}$,
$V_m, m=1,\ldots,M$ as the quadratic forms with respect to the source measures $P^{(m)}$, and $V$
as a combination of $V_m$. 

\begin{assumption}\label{A1} (Uniform equivalence)
There exists a constant $b\ge1$ such that for any $1\le m\le M$ and $f\in\cH$,
\[
b^{-1}V_0(f,f)\le V_m(f,f) \le b V_0(f,f).
\]
\end{assumption}
\begin{assumption}\label{A2} (Simultaneous diagonalization)
There are basis functions $\{\varphi_\nu\}_{\nu\ge1}\subset\cH$ and a non-decreasing positive sequence $\{\rho_\nu\}_{\nu\ge1}$ such that
\[
V_0(\varphi_\nu,\varphi_\mu)=\delta_{\nu\mu},\,\,\,\,\langle\varphi_\nu,\varphi_\mu\rangle_\cH=\rho_{\nu}\delta_{\nu\mu},\,\,\,\,\nu,\mu\ge1,
\]
where $\delta_{\nu\mu}$ is Kronecker's notation. Moreover, $\varphi_\nu$'s are uniformly bounded over $\cX$.
\end{assumption}
\begin{assumption}\label{A3} (Moment condition)
For $m=0,\ldots,M$, there exists constant $C_m>0$ such that 
\[
\textrm{$C_m^{-1}\le \E[|e^{(m)}|^2 \mid X^{(m)}]\le C_m$ and $\E[|e^{(m)}|^4 \mid X^{(m)}]\le C_m$ almost surely.}
\]
\end{assumption}
\begin{remark}
Assumption \ref{A1} indicates that the quadratic forms $V_0,\ldots,V_m$ are pairwise equivalent.
When $\cX$ is a compact subset of $\bbR^d$ and the probability densities $p^{(m)}$ associated with $P^{(m)}$
are bounded away from zero and infinity,
Assumption \ref{A1} naturally holds.
When $\cX$ is unbounded such as $\cX=\bbR^d$, Assumption \ref{A1} holds when 
$p^{(m)}(x)=p^{(0)}(x)$ for $|x|>B$, and $p^{(m)}(x)$'s are bounded away from zero and infinity
over $|x|\le B$, where $B>0$ is some constant. 
Assumption \ref{A2} is standard in smoothing spline literature which says that $V_0$ and the $\cH$-inner product can be simultaneously diagonalized by a sequence of uniformly bounded eigenfunctions $\varphi_\nu$'s associated with a sequence of eigenvalues $\rho_\nu$;
see \cite{gu1993aos}, \cite{gu2013}, \cite{shang2010ejs} and \cite{shang2020colt}. The eigenpairs $(\varphi_\nu,\rho_\nu)$'s can be found by the well-known variational method 
described in \cite{weinberg1974}; see Theorem 3.1 therein. 
Assumption \ref{A3} requires the conditional variance of $e^{(m)}$ given $X^{(m)}$
is bounded away from zero and infinity, and the fourth-order conditional moment of $e^{(m)}$ given $X^{(m)}$ is almost surely bounded.
\end{remark}

Next, we establish an RKHS framework to study the asymptotic properties for $\widehat{f}^{\textrm{cs},\omega}_{n\lambda}$ and $\widehat{f}^{\textrm{ds},\omega}_{n\lambda}$.
Consider the following inner products on $\cH$: for $f,g\in\cH$,
\begin{eqnarray*}
\langle f,g\rangle&=&V(f,g)+\lambda\langle f,g\rangle_\cH,\\
\langle f,g\rangle_m&=&V_m(f,g)+\lambda\langle f,g\rangle_\cH,\,\,m=0,\ldots,M.
\end{eqnarray*}
Let $\|\cdot\|$ and $\|\cdot\|_m$ denote the norms corresponding to $\langle\cdot,\cdot\rangle$ and $\langle\cdot,\cdot\rangle_m$.
By Assumptions \ref{A1} and \ref{A2}, we have the following obvious inequality: for any $f\in\cH$,
\begin{equation}\label{norm:equiv:cs}
b^{-2}\lambda\|f\|_\cH^2\le b^{-2}\|f\|_m^2\le\|f\|^2\le b^2\|f\|_m^2\le b^3\|f\|_0^2\le b^3(\lambda+1/\rho_1)\|f\|_\cH^2.
\end{equation}
The second to the fourth ``$\le$''s follow from $b^{-2}V_m\le V_{m'}\le b^2V_m$ and $V_m\le bV_0$, for any $m,m'=0,\ldots,M$
(by Assumption \ref{A1});
the last ``$\le$'' follows from the following fact: for any $f=\sum_{\nu\ge1}f_\nu\varphi_\nu\in\cH$,
\[
\|f\|_0^2=\sum_{\nu\ge1}f_\nu^2(1+\lambda\rho_\nu)\le\sum_{\nu\ge1}f_\nu^2(\rho_\nu/\rho_1+\lambda\rho_\nu)=(\lambda+1/\rho_1)\sum_{\nu\ge1}f_\nu^2\rho_\nu=(\lambda+1/\rho_1)\|f\|_\cH^2.
\]
Hence, $\langle\cdot,\cdot\rangle$, $\langle\cdot,\cdot\rangle_m$ and $\langle\cdot,\cdot\rangle_\cH$
are all equivalent, implying that $\cH$ under the first two are also RKHS's with reproducing kernels denoted by $K$ and $K^{(m)}$,
respectively. Let $\cP_\lambda, \cP^{(m)}_\lambda$ be bounded linear operators from $\cH$ to $\cH$ such that
\[
\langle\cP_\lambda f,g\rangle=\langle\cP^{(m)}_\lambda f,g\rangle_m=\lambda\langle f,g\rangle_\cH,\,\,\,\,\textrm{for any $f,g\in\cH$.}
\]
Existence of such operators follows from the Riesz representation theorem (see Section 2.2 of \cite{shang2013aos}).
Moreover, by Assumption \ref{A2}, there exists a universal constant $c_0$ such that
$\|K^{(0)}_x\|_0\le c_0 h^{-1/2}$ for any $x\in\cX$, where $h$ is known as the effective dimension defined by
\begin{equation}\label{def:h}
h^{-1}:=\sum_{\nu\ge1}(1+\lambda\rho_\nu)^{-1};
\end{equation}
see Section 2 of \cite{shang2020colt} for a proof.
Hence, by the Cauchy-Schwartz inequality and (\ref{norm:equiv:cs}), we have
\begin{equation}\label{bound:K_x}
\|K_x\|^2=K(x,x)=\langle K_x,K^{(0)}_x\rangle_0\le\|K_x\|_0\|K^{(0)}_x\|_0\le b\|K_x\|\times c_0h^{-1/2},
\end{equation}
implying $\|K_x\|\le b c_0 h^{-1/2}\equiv c_1 h^{-1/2}$.
Similarly, for $1\le m\le M$ and any $x\in\cX$, we have
\[
\|K^{(m)}_x\|_m^2=K^{(m)}(x,x)=\langle K^{(m)}_x,K^{(0)}_x\rangle_0\le\|K^{(m)}_x\|_0\|K^{(0)}_x\|_0\le b\|K^{(m)}_x\|_m\times c_0h^{-1/2},
\]
implying that $\|K^{(m)}_x\|_m\le c_1h^{-1/2}$.

\section{Asymptotic analysis}\label{sec:asymptotic}
In this section, we provide asymptotic analysis for the proposed estimation and inference procedures.
In Section \ref{sec:rate}, we  provide the rate of convergence and a higher-order representation for the estimators $\widehat{f}^{\textrm{cs},\omega}_{n\lambda}$ and $\widehat{f}^{\textrm{ds},\omega}_{n\lambda}$. 
In Section \ref{sec:inference}, we derive local and global bootstrap consistency based on these higher-order representations, 
which will be used to show that the bootstrap procedures developed in Section \ref{sec:inference:procedures} are statistically valid.
To facilitate asymptotic analysis, we assume that the weights $\omega^{(m)}$'s in (2) of Section \ref{sec:cs} are bounded.
The results could be extended to unbounded settings if we assume that the weights satisfy suitable tail conditions.

\subsection{Convergence rate and higher-order representations}\label{sec:rate}

Let $\cF=\{f\in\cH: \|f\|\le c_1^{-1}h^{1/2}\}$ and 
\[
a_{\lambda}:=\int_0^1\psi_2^{-1}(D(\cF,\varepsilon,\|\cdot\|_{\infty}))d\varepsilon,
\]
where $\psi_2(x)=e^{x^2}-1$ and $D(\cF,\varepsilon,\|\cdot\|_{\infty})$ is the $\varepsilon$-covering number of $\cF$ with respect to supremum norm $\|\cdot\|_\infty$. Recall $n=\sum_{m=0}^M n_m$.
\begin{theorem}\label{rate:cs}    
Suppose that Assumptions \ref{A1}, \ref{A2}, \ref{A3} hold and $a_\lambda=o(\sqrt{nh^2})$.
Then we have, as $n\to\infty$, 
\begin{eqnarray}
\|\widehat{f}^{\textrm{cs},\omega}_{n\lambda}-f^{(0)}_*\|&=&O_\Pr\left(\frac{1}{\sqrt{nh}}+\sqrt{\lambda}\right),\label{eqn:rate:cs}\\
\|\widehat{f}^{\textrm{cs},\omega}_{n\lambda}-f^{(0)}_*+S^{\textrm{cs},\omega}_{n\lambda}(f^{(0)}_*)\|&=&O_\Pr\left(\frac{a_\lambda}{\sqrt{nh^2}}\left(\frac{1}{\sqrt{nh}}+\sqrt{\lambda}\right)\right),\label{eqn:hor:cs}
\end{eqnarray}
where
\[
S^{\textrm{cs},\omega}_{n\lambda}(f^{(0)}_*)=-\frac{1}{n}\sum_{m=0}^M\sum_{i=1}^{n_m}\omega_i^{(m)}e_i^{(m)}K_{X_i^{(m)}}+\cP_\lambda f^{(0)}_*.
\]
\end{theorem}

\begin{theorem}\label{rate:ds}   
Suppose that Assumption \ref{A1}, \ref{A2}, \ref{A3} hold, and the following rate conditions are satisfied:
for $m=0,\ldots,M$,
\begin{equation}\label{rate:cond:ds}
\frac{1}{\sqrt{n_mh}}+\sqrt{\lambda}=O(1),\,\,a_\lambda=o(\sqrt{n_mh^2}),\,\,\frac{a_\lambda}{\sqrt{n_mh^2}}\left(\frac{1}{\sqrt{n_mh}}+\sqrt{\lambda}\right)=O(n^{-1/2}).
\end{equation}
Then we have, as $n\to\infty$, 
\begin{eqnarray}
\|\widehat{f}^{\textrm{ds},\omega}_{n\lambda}-f^{(0)}_*\|&=&O_\Pr((nh)^{-1/2}+\sqrt{\lambda}),\label{eqn:rate:ds}\\
\|\widehat{f}^{\textrm{ds},\omega}_{n\lambda}-f^{(0)}_*+S^{\textrm{ds},\omega}_{n\lambda}(f^{(0)}_*)\|&=&O_\Pr\left(\frac{a_\lambda}{\sqrt{nh^2}}\left(\frac{1}{\sqrt{nh}}+\sqrt{\lambda}\right)\right),\label{eqn:hor:ds}
\end{eqnarray}
where 
\begin{eqnarray*}
S^{\textrm{ds},\omega}_{n\lambda}(f^{(0)}_*)&=&-\frac{1}{n}\sum_{m=0}^M\sum_{i=n_m+1}^{2n_m}\omega_i^{(m)}\left[(\widehat{f}^{(0)}-f^{(0)}_*)(X_i^{(m)})-(\widehat{f}^{(m)}-f^{(m)}_*)(X_i^{(m)})+e_i^{(m)}\right]K_{X_i^{(m)}}\\
&&+\cP_\lambda f^{(0)}_*.
\end{eqnarray*}
\end{theorem}

Theorems \ref{rate:cs} and \ref{rate:ds} establish the convergence rates in (\ref{eqn:rate:cs}) \& (\ref{eqn:rate:ds}), and higher-order representations (\ref{eqn:hor:cs}) \& (\ref{eqn:hor:ds}) for 
$\widehat{f}^{\textrm{cs},\omega}_{n\lambda}$ and $\widehat{f}^{\textrm{ds},\omega}_{n\lambda}$.
The higher-order representations will be useful for proving bootstrap consistency in Section \ref{sec:inference}.
Letting $\omega$ involve only ones, the same rates hold for $\widehat{f}^{\textrm{cs}}_{n\lambda}$ and $\widehat{f}^{\textrm{ds}}_{n\lambda}$. 
The rate becomes optimal if we set $(nh)^{-1}\asymp\lambda$; see examples in Section \ref{sec:example}.
In the transfer learning literature, \cite{ma2023aos} and \cite{wang2023minimaxoptimaltransferlearning}
also considered the problem of estimating the mean function and obtained a nearly optimal convergence rate (up to a logarithmic factor). In the proof of \cite{ma2023aos}, the variance of the estimator was written as a quadratic form
that was bounded by the Hanson-Wright inequality and therefore a logarithmic factor arises.
In our framework, the convergence rates in Theorems \ref{rate:cs} and \ref{rate:ds}
are exactly optimal without a logarithmic factor.
We use the contraction mapping theorem to show that the variance of the estimator can be bounded by the expected squared norm of an independent sum, which can be exactly calculated without a logarithmic factor.

\subsection{Bootstrap consistency}\label{sec:inference}
Let $\Pr_*$ be the conditional probability measure given the samples $(Y^{(m)}_i,X^{(m)}_i)$'s and $\Phi(\cdot)$ be the distribution function of a standard normal variable.
For any $x\in\cX$, let $\tau^2(x)=\E\{\tau_n^2(x)\}$, where
\[
\tau_n^2(x):=\frac{1}{n^2}\sum_{m=0}^M\sum_{i=1}^{n_m}|e_i^{(m)}|^2|K(X_i^{(m)},x)|^2.
\]
By the law of large numbers, it can be shown that, for any $x\in\cX$, $\tau_n^2(x)=(1+o_\Pr(1))\tau^2(x)$, i.e.,
$\tau_n^2(x)$ and $\tau^2(x)$ are asymptotically equivalent. The former is relevant to the bootstrap variance
of $\widehat{f}^{\textrm{cs},\omega}_{n\lambda}(x)$ and $\widehat{f}^{\textrm{ds},\omega}_{n\lambda}(x)$, 
while the latter is relevant to the
asymptotic variance of 
$\widehat{f}^{\textrm{cs}}_{n\lambda}(x)$ and $\widehat{f}^{\textrm{ds}}_{n\lambda}(x)$.
Their equivalence is crucial to establish the following local bootstrap consistency.

\begin{theorem}\label{thm:local:boot} (Local Bootstrap Consistency)
Suppose that Assumptions \ref{A1}, \ref{A2}, \ref{A3} hold. 
\begin{itemize}
    \item[(1).] If $a_\lambda=o(\sqrt{nh^2})$, as well as
\begin{equation}\label{thm:local:rate}    
\textrm{$\frac{a_\lambda}{h}\left(\frac{1}{\sqrt{nh}}+\sqrt{\lambda}\right)=o(1)$ and $\sqrt{nh}(\cP_\lambda f^{(0)}_*)(x)=o(1)$,} 
\end{equation}
then for any $x\in\cX$ and $u\in\bbR$, as $n\to\infty$,
\begin{eqnarray*}
\Pr_*\left(\widehat{f}^{\textrm{cs},\omega}_{n\lambda}(x)-\widehat{f}^{\textrm{cs}}_{n\lambda}(x)\le\tau_n(x)u\right)&\overset{\Pr}{\to}&\Phi(u),\\    
\Pr\left(\widehat{f}^{\textrm{cs}}_{n\lambda}(x)-f^{(0)}_*(x)\le \tau(x)u\right)&\to&\Phi(u).
\end{eqnarray*}
    \item[(2).] If the rate conditions (\ref{rate:cond:ds}) hold, as well as (\ref{thm:local:rate}),
    then for any $x\in\cX$ and $u\in\bbR$, as $n\to\infty$,
\begin{eqnarray*}
\Pr_*\left(\widehat{f}^{\textrm{ds},\omega}_{n\lambda}(x)-\widehat{f}^{\textrm{ds}}_{n\lambda}(x)\le\tau_n(x)u\right)&\overset{\Pr}{\to}&\Phi(u),\\    
\Pr\left(\widehat{f}^{\textrm{ds}}_{n\lambda}(x)-f^{(0)}_*(x)\le \tau(x)u\right)&\to&\Phi(u).
\end{eqnarray*}
\end{itemize}
\end{theorem}

\begin{theorem}\label{thm:global:boot} (Global Bootstrap Consistency)
Suppose that Assumptions \ref{A1}, \ref{A2}, \ref{A3} hold and $e^{(m)}$'s additionally satisfy
$\E[\exp\left(\kappa|e^{(m)}|^2\right)]<\infty$ for some constant $\kappa>1$. 
\begin{itemize}
    \item[(1).] If $a_\lambda=o(\sqrt{nh^2})$, as well as
    \begin{equation}\label{thm:glb:rate}
\textrm{$nh^2\gg1$, $nh\gg(\log{n})^2$, $\frac{a_\lambda}{h}\left(\frac{1}{\sqrt{nh}}+\sqrt{\lambda}\right)=o(1)$, $nV(\cP_\lambda f^{(0)}_*)=o(1)$,}
    \end{equation} 
    then there exist positive random variables $\sigma_n^2,\zeta_n^2$ determined by the data, and positive scalars $\sigma^2,\zeta^2$ with $\sigma_n^2=(1+o_\Pr(1))\sigma^2$ and $\zeta_n^2=(1+o_\Pr(1))\zeta^2$, such that for any $u\in\bbR$, as $n\to\infty$
\begin{eqnarray*}
\Pr_*\left(V(\widehat{f}^{\textrm{cs},\omega}_{n\lambda}-\widehat{f}^{\textrm{cs}}_{n\lambda})\le\sigma_n^2+u\zeta_n\right)
&\overset{\Pr}{\to}&\Phi(u),\\
\Pr\left(V(\widehat{f}^{\textrm{cs}}_{n\lambda}-f^{(0)}_*)\le\sigma^2+u\zeta\right)&\to&\Phi(u).
\end{eqnarray*}
    
    \item[(2).] If the rate conditions (\ref{rate:cond:ds}) hold, as well as (\ref{thm:glb:rate}),
    then there exist positive random variables $\sigma_n^2,\zeta_n^2$ determined by the data, and positive scalars $\sigma^2,\zeta^2$ with $\sigma_n^2=(1+o_\Pr(1))\sigma^2$ and $\zeta_n^2=(1+o_\Pr(1))\zeta^2$, such that for any $u\in\bbR$, as $n\to\infty$
\begin{eqnarray*}
\Pr_*\left(V(\widehat{f}^{\textrm{ds},\omega}_{n\lambda}-\widehat{f}^{\textrm{ds}}_{n\lambda})\le\sigma_n^2+u\zeta_n\right)
&\overset{\Pr}{\to}&\Phi(u),\\
\Pr\left(V(\widehat{f}^{\textrm{ds}}_{n\lambda}-f^{(0)}_*)\le\sigma^2+u\zeta\right)&\to&\Phi(u).
\end{eqnarray*}
\end{itemize}
\end{theorem}

Theorems \ref{thm:local:boot} and \ref{thm:global:boot} establish bootstrap consistency,
that is, the conditional distribution of $\widehat{f}^{\textrm{cs},\omega}_{n\lambda}-\widehat{f}^{\textrm{cs}}_{n\lambda}$
(or $\widehat{f}^{\textrm{ds},\omega}_{n\lambda}-\widehat{f}^{\textrm{ds}}_{n\lambda}$) given the data is asymptotically equivalent to
the distribution of $\widehat{f}^{\textrm{cs}}_{n\lambda}-f^{(0)}_*$ (or $\widehat{f}^{\textrm{ds}}_{n\lambda}-f^{(0)}_*$).
The proofs rely on the higher-order representations for 
$\widehat{f}^{\textrm{cs},\omega}_{n\lambda}$ and $\widehat{f}^{\textrm{ds},\omega}_{n\lambda}$ established in Theorems \ref{rate:cs} and \ref{rate:ds}, and the fact that the quadratic forms associated with these higher-order representations are asymptotically equally distributed. 
In the literature, the asymptotic distributions of the quadratic forms are typically established by the central limit theorem of \cite{de1987central}
(see \citealt{shang2013aos, shang2015aos, shang2020colt}). However, this strategy is no longer applicable here since the probability measure $\Pr_*$ is conditional and the terms in the higher-order representations are 
dependent, as they all depend on the initial estimators of the mean functions $\widehat{f}^{(0)}, \ldots, \widehat{f}^{(M)}$.
To address this, we use a conditional version of \cite{de1987central} to address these challenges.

Theorems \ref{thm:local:boot} and \ref{thm:global:boot} are useful in justifying the validity of the pointwise bootstrap confidence interval for $f^{(0)}_*(x)$ at any $x\in\cX$, and the global bootstrap confidence region for $f^{(0)}_*$, as proposed in Section \ref{sec:inference:procedures}.
Therefore, one can use bootstrap percentiles to approximate the width of the confidence interval or radius of the confidence region,
which can be easily implemented in practice. 
See Remark \ref{rem:jusfity} for more details.
For technical convenience, the expressions of $\sigma_n^2,\zeta_n^2,\sigma^2,\zeta^2$ in Part (2) of Theorem \ref{thm:global:boot} are not explicitly provided, which can be found in the proof. 
There is rich literature on bootstrap methods in parametric settings, and so we only list some classic textbooks
(see \citealt{hall1992bootstrap,efron1994introduction,davison1997bootstrap,good2013resampling,shao1995jackknife,givens2012computational,chernick2008bootstrap}). In nonparametric settings, 
bootstrap methods based on resampled data or multiplier weights 
have been employed in nonparametric regression or density estimation
problems (see \citealt{hall2001aos,hall2013aos,calonico2018jasa,spokoiny2015aos,cheng2019ejs,KATO2018joe}).
Note that Theorems \ref{thm:local:boot} and \ref{thm:global:boot} belong to the multiplier bootstrap category
established under the present data integration architecture, which are new in the literature.

\begin{remark}\label{rem:jusfity}
Theorems \ref{thm:local:boot} and \ref{thm:global:boot} can justify the validity of our methods.
For instance, Theorem \ref{thm:local:boot} says that we can construct an asymptotic $100(1-\alpha)\%$ CI for $f^{(0)}_*(x)$ as
$(\widehat{f}^{\textrm{cs}}_{n\lambda}(x)-\tau(x)z_{1-\frac{\alpha}{2}},\widehat{f}^{\textrm{cs}}_{n\lambda}(x)-\tau(x)z_{\frac{\alpha}{2}})$, where $z_{\frac{\alpha}{2}}$ and $z_{1-\frac{\alpha}{2}}$ are $100(\frac{\alpha}{2})\%$- and  $100(1-\frac{\alpha}{2})\%$-quantiles of the standard normal distribution.
Since $\tau_n(x)$ is close to $\tau(x)$ with probability approaching one,
and $\tau_n(x)z_{\frac{\alpha}{2}}$ and $\tau_n(x)z_{1-\frac{\alpha}{2}}$
are $100(\frac{\alpha}{2})\%$- and  $100(1-\frac{\alpha}{2})\%$-quantiles of $\widehat{f}^{\textrm{cs},\omega}_{n\lambda}(x)-\widehat{f}^{\textrm{cs}}_{n\lambda}(x)$ conditional on the data,
we can use the empirical bootstrap percentiles to estimate $\tau(x)z_{\frac{\alpha}{2}}$ and $\tau(x)z_{1-\frac{\alpha}{2}}$.
This justifies the asymptotic validity of $\textrm{CI}^{\textrm{cs}}_\alpha(x)$.
The asymptotic validity of $\textrm{CI}^{\textrm{ds}}_\alpha(x)$, $R^{\textrm{cs}}_\alpha$,
$R^{\textrm{ds}}_\alpha$ can be similarly justified using Theorem \ref{thm:local:boot} or \ref{thm:global:boot}.
\end{remark}

\subsection{Examples}\label{sec:example}
Let $N=n_1+\cdots+n_M$ be the total number of source samples. 
We will use Theorems \ref{rate:cs}-\ref{thm:global:boot} to investigate a quantitative relationship between $n_0$ and $N$ such that 
$\widehat{f}^{\textrm{cs},\omega}_{n\lambda}$ and $\widehat{f}^{\textrm{ds},\omega}_{n\lambda}$ achieve satisfactory properties, i.e., the optimal convergence rate and bootstrap consistency.
We provide two examples in which the eigenvalues have polynomial and exponential orders, respectively.
\begin{example} \label{exa:polynomial}
\normalfont
Suppose $\cX=[0,1]^d$ and $\cH=\otimes^d H^\beta$ is the completed tensor product space of $H^\beta$ with
itself $d$ times, where $H^\beta$ is the $\beta$-order periodic spline space on $[0,1]$ with $\beta>\frac{3+\sqrt{5}}{4}d\approx 1.309d$.
The RKHS norm $\|\cdot\|_\cH$ of $\cH$ is naturally induced by the Soblev norm of $H^\beta$.
It is known that $\rho_\nu\asymp \nu^{2\beta/d}$; see Lemma 3.1 of \cite{xing2024minimax}.
Normally, we need $\beta>d/2$ to make $\cH$ an RKHS \citep{w90}, 
but here we need $\beta$ to be larger to ensure that the theorem conditions hold.
By (\ref{def:h}) it can be checked that $\lambda=h^{2\beta/d}$.
Let $\cF_0=\{f\in\cH: \|f\|_\infty\le1, \|f\|_\cH\le 1\}$.
Then $\cF\subset c_1\sqrt{h/\lambda}\cF_0$, where $c_1 = bc_0$ is the constant specified in Section \ref{sec:assumptions}.
By Example 3 of Section 6 in \cite{cucker2001mathfoundation}, we have 
\begin{equation}\label{computing:covering:number}
\log{D(\varepsilon,\cF,\|\cdot\|_\infty)}\le\log{D(\varepsilon,c_1\sqrt{h/\lambda}\cF_0,\|\cdot\|_\infty)}
\lesssim \left(\frac{c_1\sqrt{h/\lambda}}{\varepsilon}\right)^{\frac{d}{\beta}}.
\end{equation}
So $a_\lambda\lesssim (h/\lambda)^{\frac{d}{4\beta}}$. For simplicity, we only consider a special choice of $\lambda,h$:
\begin{equation}\label{example:1:h}
\textrm{$(nh)^{-1}\asymp\lambda$ so that $h\asymp n^{-\frac{d}{2\beta+d}}$ and $\lambda\asymp n^{-\frac{2\beta}{2\beta+d}}$.}
\end{equation}
Under (\ref{example:1:h}), we have $a_\lambda\lesssim n^{\frac{d(2\beta-d)}{4\beta(2\beta+d)}}=o(\sqrt{nh^2})$.
Hence, by Theorem \ref{rate:cs}, $\|\widehat{f}^{\textrm{cs},\omega}_{n\lambda}-f^{(0)}_*\|=O_\Pr(n^{-\frac{\beta}{2\beta+d}})$ 
which is optimal.
Moreover, we assume the following relationship between $n_m$ and $n$:
\begin{equation}\label{example1:eqn1}
n^{\frac{4\beta^2+10\beta d-d^2}{4\beta(2\beta+d)}}\ll n_m\le n, m=0,\ldots,M.
\end{equation}
(Note that $\frac{4\beta^2+10\beta d-d^2}{4\beta(2\beta+d)}<1$ due to $\beta>\frac{3+\sqrt{5}}{4}d$, 
so (\ref{example1:eqn1}) makes sense.)
It can be directly checked that (\ref{example1:eqn1}) implies (\ref{rate:cond:ds}).
By Theorem \ref{rate:ds}, 
$\|\widehat{f}^{\textrm{ds},\omega}_{n\lambda}-f^{(0)}_*\|=O_\Pr(n^{-\frac{\beta}{2\beta+d}})$ which achieves minimax optimality.

We also consider the bias term $(\cP_\lambda f^{(0)}_*)(x)$.
Before that, we follow \cite{shang2013aos} or \cite{shang:ejs:2020} to write the reproducing kernel $K$ in an alternative form:
\[
\textrm{$K(x,x')=\sum_\nu\frac{\phi_\nu(x)\phi_\nu(x')}{1+\lambda\theta_\nu}$, $x,x'\in\cX$,}
\]
where $(\phi_\nu,\theta_\nu)$'s are eigen-pairs such that $V(\phi_\nu,\phi_\mu)=\delta_{\nu\mu}$ and
$\langle\phi_\nu,\phi_\mu\rangle_\cH=\theta_\nu\delta_{\nu\mu}$.
By Assumption \ref{A1}, $V$ and $V_0$ are equivalent. 
One can think of both eigenvalues $\theta_\nu$'s and $\rho_\nu$'s being generated using the variational method described in \cite{weinberg1974}.
By the mapping principle in Section 3 therein, we have that $\theta_\nu$'s and $\rho_\nu$'s are equivalent,
i.e., $\theta_\nu\asymp\nu^{2\beta/d}$ as well. 
Now let us write $f^{(0)}_*=\sum_\nu V(f^{(0)}_*,\phi_\nu)\phi_\nu$ and suppose for a constant $\kappa$ with $1+\frac{d}{2\beta}<\kappa\le2$,
\begin{equation}\label{f*:0:additional:cond}
\textrm{$\sum_\nu|V(f^{(0)}_*,\phi_\nu)|^2\theta_\nu^\kappa<\infty$.}
\end{equation}
Similar to Section 5.2 of \cite{shang2013aos}, it can be shown that 
\[
\cP_\lambda f^{(0)}_*=\sum_\nu V(f^{(0)}_*,\phi_\nu)\frac{\lambda\theta_\nu}{1+\lambda\theta_\nu}\phi_\nu.
\]
Therefore,
\[
\|\cP_\lambda f^{(0)}_*\|^2=\sum_\nu|V(f^{(0)}_*,\phi_\nu)|^2\frac{(\lambda\theta_\nu)^2}{1+\lambda\theta_\nu}
\lesssim \lambda^\kappa\sum_\nu|V(f^{(0)}_*,\phi_\nu)|^2\theta_\nu^\kappa=O(\lambda^\kappa).
\]
By (\ref{bound:K_x}), we have 
\[
\textrm{$(\cP_\lambda f^{(0)}_*)(x)=O(\sqrt{\lambda^\kappa/h})$ and $V(\cP_\lambda f^{(0)}_*)=O(\lambda^\kappa)$.}
\]
By direct calculations, it can be checked that (\ref{thm:local:rate}) and (\ref{thm:glb:rate}) hold.
This implies that, under (\ref{example:1:h}) and (\ref{f*:0:additional:cond}), 
Part (1) of Theorems \ref{thm:local:boot}-\ref{thm:global:boot} hold;
under (\ref{example1:eqn1}) and (\ref{f*:0:additional:cond}), 
Part (2) of Theorems \ref{thm:local:boot}-\ref{thm:global:boot} hold.

As a side remark, (\ref{example1:eqn1}) indicates an interesting relationship between $n_0$ and $N$. 
Suppose $n_0=o(N)$, i.e., the size of target sample is far less than the size of source sample. Then (\ref{example1:eqn1}) together with $n\asymp N$ implies that
\begin{equation}\label{poly:decay:exam:relationship:n_0:N}
N^{\frac{4\beta^2+10\beta d-d^2}{4\beta(2\beta+d)}}\ll n_0\ll N.
\end{equation}
This means that $N$ is not arbitrarily large (compared to $n_0$) to guarantee the optimality of $\widehat{f}^{\textrm{ds},\omega}_{n\lambda}$ as well as bootstrap consistency. We ignore the case $N=O(n_0)$ since otherwise the optimal rate $n^{-\frac{\beta}{2\beta+d}}$
is equivalent to $n_0^{-\frac{\beta}{2\beta+d}}$, i.e.,
the source sample does not enhance the optimal convergence rate. 
\end{example}

\begin{remark}
In the transfer learning literature, the relationship between $n_m$ and $N$ is typically not required,
but the source and target distributions must be similar to guarantee transferability. 
For instance, in \cite{li2021jrssb,li2023jasa,li2024jasa,ye2023jasa,wang2023minimaxoptimaltransferlearning},
a pre-screening step is performed to select the sources that are sufficiently similar to the target,
based on which estimation of the target parameter is executed.
In the current framework, we don't require the target and source distributions to be close to each other;
literally they can be arbitrarily dissimilar.
The cost is that the sizes of the target and source samples must satisfy a relationship like (\ref{poly:decay:exam:relationship:n_0:N}) to make the approaches work. This is consistent with our intuition that data integration may underperform if the size of source data is significantly large compared to that of target data.
The relationship varies from case to case, as seen in the following example of exponential-order eigenvalues.
\end{remark}

\begin{example}\normalfont
Suppose $\cX=\bbR^d$ and $R(x,x')=e^{-(|x-x'|/\iota)^\beta}$ is an exponential-type kernel with positive constants $\beta, \iota>0$.
Let $\cH$ be the RKHS of functions on $\cX$ with reproducing kernel $R$ and naturally associated norm $\|\cdot\|_\cH$; see Chapter 7.4 of \cite{berlinet2011reproducing}.
It is well known that the eigenvalues satisfy $\rho_\nu\asymp e^{\nu^\beta}$ \citep{sollich2004springer}.
By (\ref{def:h}), it can be verified that $h\asymp (-\log\lambda)^{-\frac{1}{\beta}}$; see Example II of \cite{zhao2016aos}. 
By (\ref{computing:covering:number}) and Proposition 1 of \cite{ZHOU2002739}, we have
\[
\log{D(\varepsilon,\cF,\|\cdot\|_\infty)}\le\log{D(\varepsilon,c_1\sqrt{h/\lambda}\cF_0,\|\cdot\|_\infty)}
\lesssim \left(\log\frac{c_1\sqrt{h/\lambda}}{\varepsilon}\right)^{d+1}.
\]
So $a_\lambda\lesssim [\log(h/\lambda)]^{\frac{d+1}{2}}$. We only consider the following special choice of $\lambda,h$:
\begin{equation}\label{example:1:h:exp}
\textrm{$(nh)^{-1}\asymp\lambda$ so that $h\asymp (\log{n})^{-\frac{1}{\beta}}$ and $\lambda\asymp\frac{(\log{n})^{\frac{1}{\beta}}}{n}$.} 
\end{equation}
Under (\ref{example:1:h:exp}), we have $a_\lambda\lesssim (\log{n})^{\frac{d+1}{2}}=o(\sqrt{nh^2})$.
By Theorem \ref{rate:cs}, 
$\|\widehat{f}^{\textrm{cs},\omega}_{n\lambda}-f^{(0)}_*\|=O_\Pr(n^{-\frac{1}{2}}(\log{n})^{\frac{1}{2\beta}})$ which is optimal.
Moreover, we assume the following relationship between $n_m$ and $n$:
\begin{equation}\label{example1:eqn1:exp}
\sqrt{n}(\log{n})^{\frac{d+1}{2}+\frac{3}{2\beta}}\ll n_m\le n, m=0,\ldots,M.
\end{equation}
It can be directly checked that (\ref{example1:eqn1:exp}) implies (\ref{rate:cond:ds}).
By Theorem \ref{rate:ds}, 
$\|\widehat{f}^{\textrm{ds},\omega}_{n\lambda}-f^{(0)}_*\|=O_\Pr(n^{-\frac{1}{2}}(\log{n})^{\frac{1}{2\beta}})$ which is optimal.

Next, suppose that for a constant $\kappa>\frac{1}{\beta}$, $f^{(0)}_*$ satisfies
\begin{equation}\label{f*:0:additional:cond:exp}
\textrm{$\sum_\nu|V(f^{(0)}_*,\phi_\nu)|^2\theta_\nu(\log\theta_\nu)^\kappa<\infty$.}
\end{equation}
Then we have
\begin{eqnarray*}
\|\cP_\lambda f^{(0)}_*\|^2&=&\sum_\nu|V(f^{(0)}_*,\phi_\nu)|^2\frac{(\lambda\theta_\nu)^2}{1+\lambda\theta_\nu}\\
&=&\sum_\nu|V(f^{(0)}_*,\phi_\nu)|^2\lambda\theta_\nu(\log\theta_\nu)^\kappa\frac{\lambda\theta_\nu}{(1+\lambda\theta_\nu)(\log\theta_\nu)^\kappa}=O(\lambda(-\log\lambda)^{-\kappa}),
\end{eqnarray*}
where the last big-$O$-term follows from the following trivial calculus fact: for any $\nu\ge1$,
\[
\frac{\lambda\theta_\nu}{(1+\lambda\theta_\nu)(\log\lambda\theta_\nu-\log\lambda)^\kappa}\le 
\sup_{2\lambda\le x<\infty}\frac{x}{(1+x)(\log{x}-\log\lambda)^\kappa}\lesssim (-\log\lambda)^{-\kappa}.
\]
Then we have 
\[
nV(\cP_\lambda f^{(0)}_*)=O((\log{n})^{\frac{1}{\beta}-\kappa})=o(1),
\]
and 
\[
\sqrt{nh}|(\cP_\lambda f^{(0)}_*)(x)|\le \sqrt{nh} c_1h^{-1/2}\|\cP_\lambda f^{(0)}_*\|\lesssim (\log{n})^{\frac{1}{2\beta}-\frac{\kappa}{2}}=o(1).
\]
Therefore, it can be checked that (\ref{thm:local:rate}) and (\ref{thm:glb:rate}) hold.
This implies that, under (\ref{example:1:h:exp}) and (\ref{f*:0:additional:cond:exp}), 
Part (1) of Theorems \ref{thm:local:boot}-\ref{thm:global:boot} hold;
under (\ref{example1:eqn1:exp}) and (\ref{f*:0:additional:cond:exp}), 
Part (2) of Theorems \ref{thm:local:boot}-\ref{thm:global:boot} hold.
Finally, when $n_0=o(N)$, we notice that (\ref{example1:eqn1:exp}) implies the following relationship 
\begin{equation}\label{exp:decay:exam:relationship:n_0:N}
\sqrt{N}(\log{N})^{\frac{d+1}{2}+\frac{3}{2\beta}}\ll n_0\ll N.
\end{equation}
Interpretations of (\ref{exp:decay:exam:relationship:n_0:N}) are similar to (\ref{poly:decay:exam:relationship:n_0:N}),
hence, are omitted.
\end{example}

\section{Simulation studies}\label{sec:sim} 
In this section, we perform simulation studies to examine the finite-sample performance of our proposed estimator. In particular, we are interested in its estimation and inference performance as $N/n_0$ varies, where $N$ denotes the total sample size of the source data. For numerical implementations, as in Example \ref{exa:polynomial}, we choose $\cH=\otimes^d H^2$ as the completed tensor product space of $H^2$ with itself $d$ times, where $H^2$ is the second order periodic spline on $[0, 1]$. More details regarding this RKHS can be found in Chapters 4.2 and 4.3 of \cite{gu2013}. Moreover, we employ function {\it ssanova} in the R package gss \citep{gss} to obtain $\hat{f}^{(m)}, m = 0, \ldots, M$ and the final estimators $\widehat{f}^{\textrm{ds},\omega}_{n\lambda}$ and $\widehat{f}^{\textrm{ds}}_{n\lambda}$. We only consider the distribution shift scenario since this is more general and flexible. As suggested in \cite{gss}, we use generalized cross-validation (GCV) to choose the tuning parameter $\lambda$. 

\subsection{Simulation settings}
To examine the performance of $\widehat{f}^{\textrm{ds},\omega}_{n\lambda}$ and the proposed inference tool as $N/n_0$ varies, we consider the following two settings, where only a single source dataset is considered, i.e., $N = n_1$, and $d = 1$ and 2, respectively. 

{\bf Setting 1}: For $i = 1, \ldots, n_m$, a univariate covariate $x_i^{(m)}$ is independently sampled from $U(0, 1)$ for $m = 0, 1$. Then we independently generate the response as $y_i^{(m)} = f^{(m)}_*(x_i^{(m)}) + \epsilon_i^{(m)}$, where
$$
f^{(m)}_*(x_i^{(m)}) = 3\sin\left[2\pi \left(x_i^{(m)} - \tau^{(m)}\right)\right] - \exp(x_i^{(m)}) + \{x_i^{(m)}\}^2
$$
with $\tau^{(0)} = 0$ and $\tau^{(1)} \in \{0.05, 0.10, 0.15\}$, and $\epsilon_i^{(m)} \sim N(0, \sigma^2)$. A proper $\sigma^2$ is chosen such that the signal-to-noise ratio, $\var\{f^{(0)}_*(x_i^{(0)})\}/\sigma^2$, is equal to 10. Obviously, $\tau^{(m)}$ measures the similarity between the source data and the target data. 

{\bf Setting 2}: For $i = 1, \ldots, n_m$, we consider a bivariate covariate $\bx_i^{(m)} = (x_{i1}^{(m)}, x_{i2}^{(m)})^{\T}$, where each component is independently sampled from $U(0, 1)$ for $m = 0, 1$. Then we independently generate the response as $y_i^{(m)} = f^{(m)}_*(\bx_i^{(m)}) + \epsilon_i^{(m)}$, where
$$
f^{(m)}_*(\bx_i^{(m)}) =  3\sin\left[2\pi \left(x_{i1}^{(m)} - \tau^{(m)}\right)\right] - \exp(x_{i1}^{(m)}) + \{x_{i1}^{(m)} - x_{i2}^{(m)}\}^2
$$
with $\tau^{(0)} = 0$ and $\tau^{(1)} \in \{0.05, 0.10, 0.15\}$, and $\epsilon_i^{(m)} \sim N(0, \sigma^2)$. We choose $\sigma^2$ with the same method as in Setting 1. 

In both settings, we set $n_0 = 200$ or $1,\!000$ and various $n_1/n_0$ from $\{0, 0.25, 0.5, 1, 2, 4, 8\}$ to investigate the effect of the size of the target data as well as the source-to-target ratio. In particular, $n_1/n_0=0$ means that only target data are used to estimate $f^{(0)}_*$.
$10,\!000$ independent Monte Carlo simulation runs are carried out for each of the above scenarios. 

\subsection{Simulation results}
\begin{figure}[ht]
	\centering
	{\includegraphics[width=14cm]{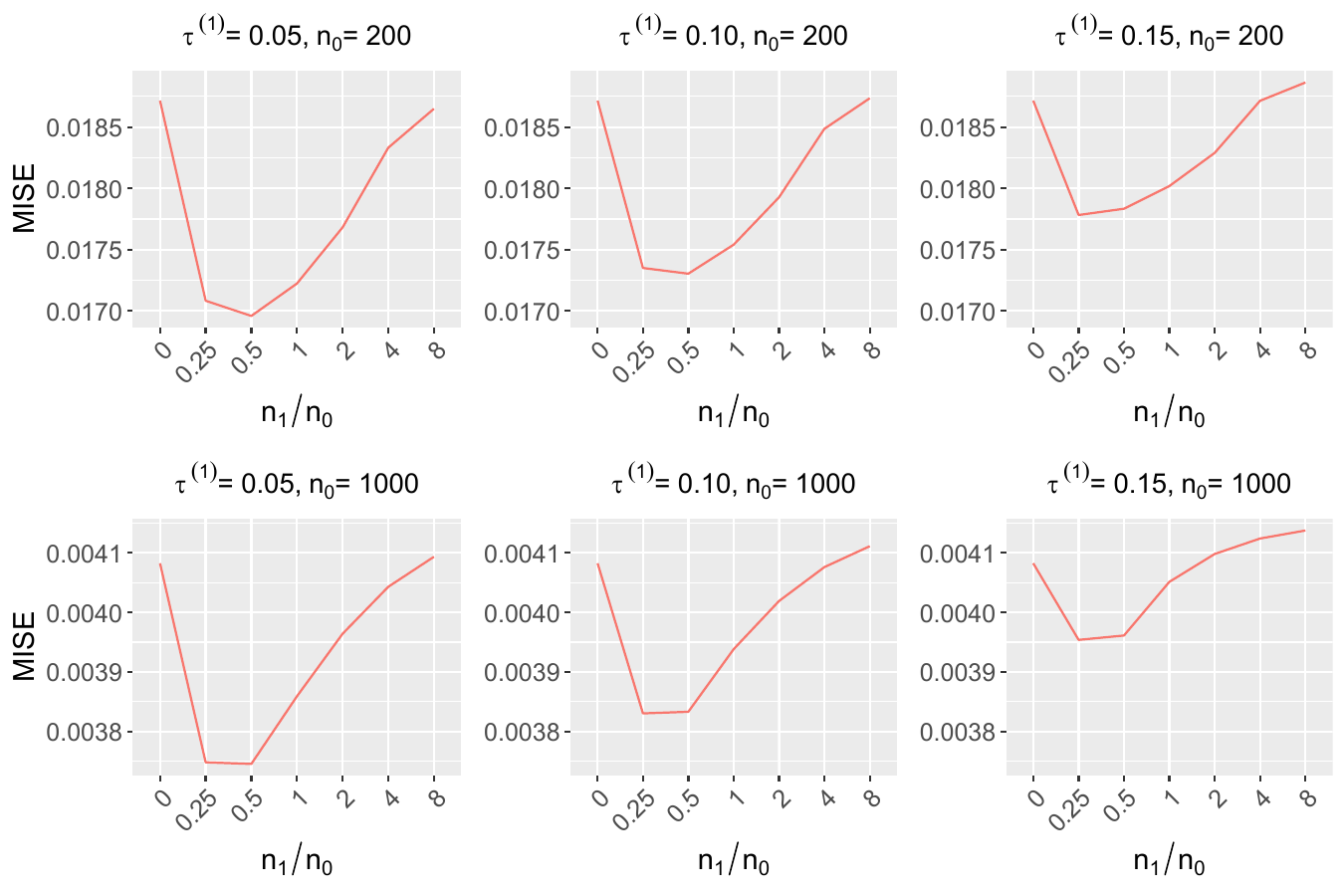}}
	\caption{
Simulation results of MISEs for $\widehat{f}^{\textrm{ds}}_{n\lambda}$
 across $10,\!000$ Monte Carlo replicates in Setting 1. The sample sizes of the target data for the top and bottom panels are $n_0 = 200$ and $1,\!000$, respectively. The similarity index $\tau^{(1)}$ is equal to 0.05, 0.10 and 0.15 from the left to the right panels.}
	\label{fig:est-example1}
\end{figure}

\begin{figure}[ht]
	\centering
	{\includegraphics[width=14cm]{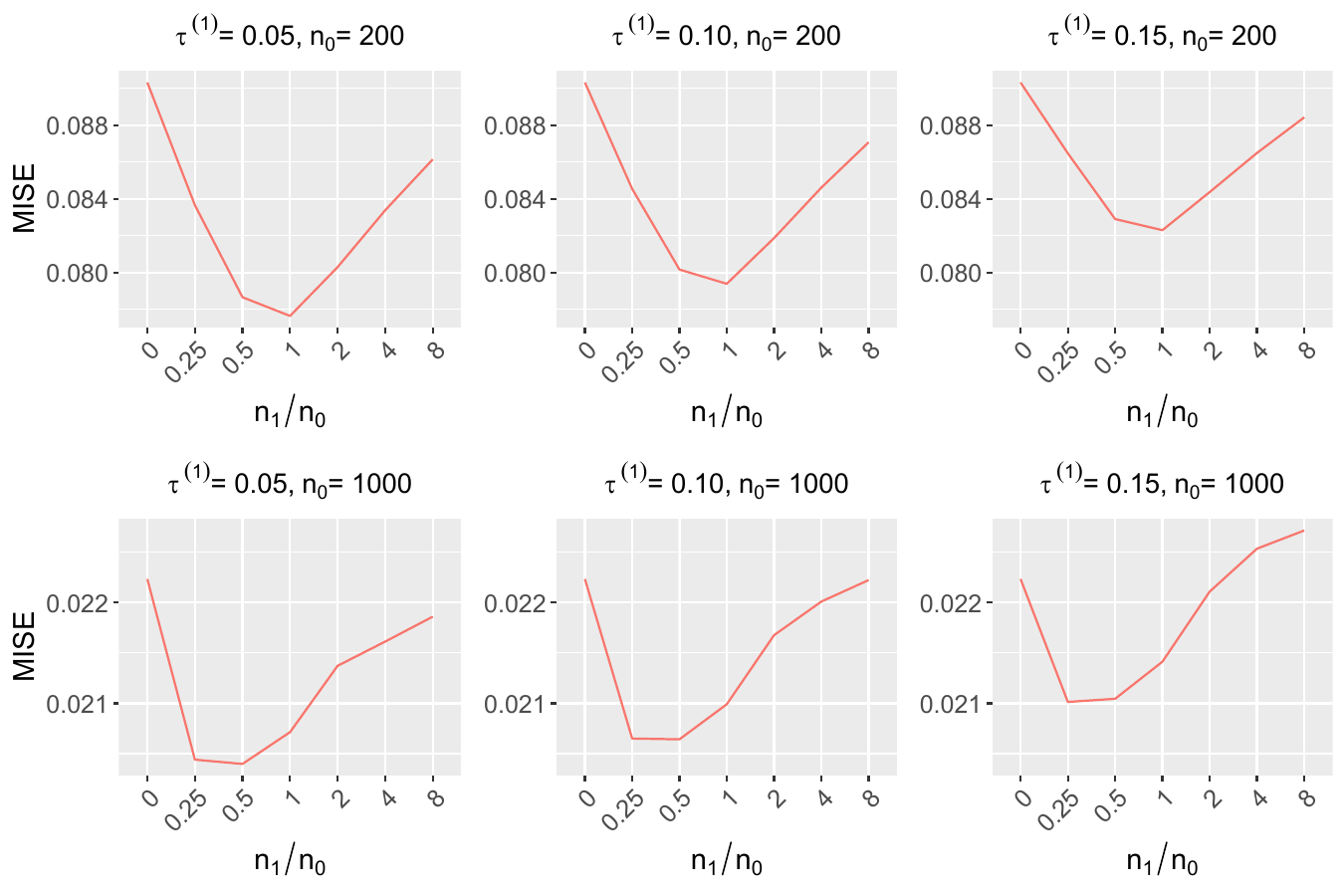}}
	\caption{
Simulation results of MISEs for $\widehat{f}^{\textrm{ds}}_{n\lambda}$
 across $10,\!000$ Monte Carlo replicates in Setting 2. The sample sizes of the target data for the top and bottom panels are $n_0 = 200$ and $1,\!000$, respectively. The similarity index $\tau^{(1)}$ is equal to 0.05, 0.10 and 0.15 from the left to the right panels.}
	\label{fig:est-example2}
\end{figure}

To evaluate the performance of $\widehat{f}^{\textrm{ds}}_{n\lambda}$ defined in Section \ref{sec:methods}, we consider the integrated squared error (ISE), which is defined as $\text{ISE}(\widehat{f}^{\textrm{ds}}_{n\lambda}) = \int_{[0, 1]^d} \{\widehat{f}^{\textrm{ds}}_{n\lambda}(x) - f^{(0)}_*(x)\}^2 dx$. Figures \ref{fig:est-example1} and \ref{fig:est-example2} display the mean ISEs (MISE) of $\widehat{f}^{\textrm{ds}}_{n\lambda}$ across $10,\!000$ Monte Carlo simulations as $n_1/n_0$ varies in Settings 1 and 2, respectively. Both figures demonstrate that as $n_1/n_0$ increases from zero to 0.5 (0.25 or 1 in some settings), the estimator 
becomes more accurate in estimating $f^{(0)}_*$; when $n_1/n_0$ increases from around 0.5 to 8, the estimation performance becomes worse.
Moreover, as $n_0$ increases, the estimator becomes more accurate when $n_1/n_0$ does not change. The U-shapes displayed in all panels in these two figures justify the necessity of the assumption imposed on the order of $n_0$ relative to $n_1$.
We also observe that, holding $n_1/n_0$ constant, the improvement in estimation performance decreases as the similarity index $\tau^{(1)}$ increases.
Last but not least, Figures \ref{fig:est-example1} and \ref{fig:est-example2}  present the results for $n_1 / n_0 = 0$, i.e., only the target data are used to estimate the true regression function. These results demonstrate the effectiveness of our method in estimation via data integration.

\begin{figure}[ht]
	\centering
	{\includegraphics[width=\textwidth]{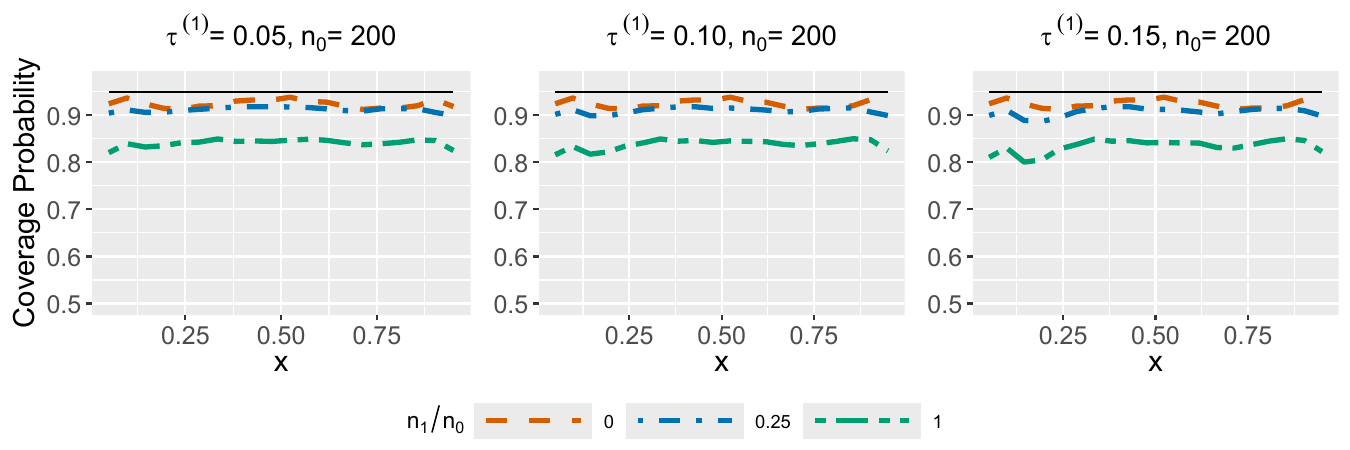}}
	\caption{ Coverage probabilities of the 95\% pointwise confidence intervals $\textrm{CI}^{\textrm{ds}}_\alpha(x)$
	proposed in Section \ref{sec:inference:procedures} for Setting 1 with $n_0 = 200$ over $10,\!000$ Monte Carlo simulations. From the left to the right panels , $\tau^{(1)}$ is equal to 0.05, 0.10 and 0.15, respectively. In all panels, the solid black line represents the nominal coverage probability 95\%, while the dashed red, dotted blue and dot dashed green lines represent the true coverage probabilities corresponding to $n_1/n_0$ equal to 0, 0.25 and 1, respectively. }
	\label{fig:CP100-example1}
\end{figure}

\begin{figure}[ht]
	\centering
	{\includegraphics[width=\textwidth]{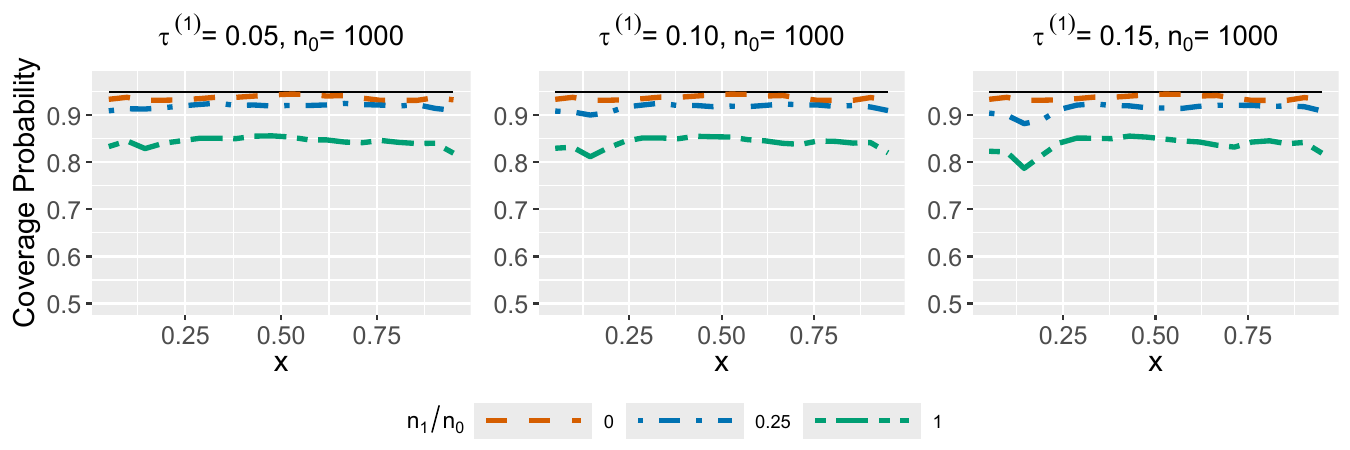}}
	\caption{ Coverage probabilities of the 95\% pointwise confidence intervals $\textrm{CI}^{\textrm{ds}}_\alpha(x)$
	proposed in Section \ref{sec:inference:procedures} for Setting 1 with $n_0 = 1,\!000$ over $10,\!000$ Monte Carlo simulations. From the left to the right panels, $\tau^{(1)}$ is equal to 0.05, 0.10 and 0.15, respectively. In all panels, the solid black line represents the nominal coverage probability 95\%, while the dashed red, dotted blue and dot dashed green lines represent the true coverage probabilities corresponding to $n_1/n_0$ equal to 0, 0.25 and 1, respectively. }
	\label{fig:CP500-example1}
\end{figure}

Figures \ref{fig:CP100-example1} and \ref{fig:CP500-example1} showcase the coverage probabilities of the pointwise confidence intervals
$\textrm{CI}^{\textrm{ds}}_\alpha(x)$ proposed in Section \ref{sec:inference:procedures}. The results are based on $10,\!000$ Monte Carlo simulations under Setting 1 with respect to $n_0 = 200$ and $1,\!000$, respectively. The six panels in these two figures display similar patterns; the coverage probabilities of the confidence intervals with a relatively smaller $n_1/n_0$ are closer to the nominal level, compared with those with a large $n_1/n_0$. This indicates the necessity of the assumptions on the order of $n_0$ relative to $n_1$ to guarantee satisfactory coverage probabilities.


\section{Real data analysis}\label{sec:realdata}

We illustrate our method using an example from the ADNI dataset to investigate Alzheimer's disease by examining the association between cognitive scores and hippocampal volume. Hippocampal volume, a well-established MRI biomarker, is crucial for diagnosing brain atrophy \citep{morar2022study}. Alzheimer's disease, the most common form of dementia in the United States, predominantly affects individuals aged 65 and older where it is being referred to as late onset Alzheimer's disease (LOAD), with an estimated prevalence of 1 in 9 people \citep{reitz2020late}. However, the true prevalence and incidence rates of early onset Alzheimer's disease (EOAD)—diagnosed in those under 65—are believed to be underreported. This relative scarcity of data on EOAD calls for further research to characterize its clinical presentation. In our case study, we address this gap by employing the proposed method to perform regression analysis of cognitive scores based on hippocampal volume in EOAD, through integrating data from LOAD.

\begin{figure}[ht]
	\centering
	{\includegraphics[width=14cm]{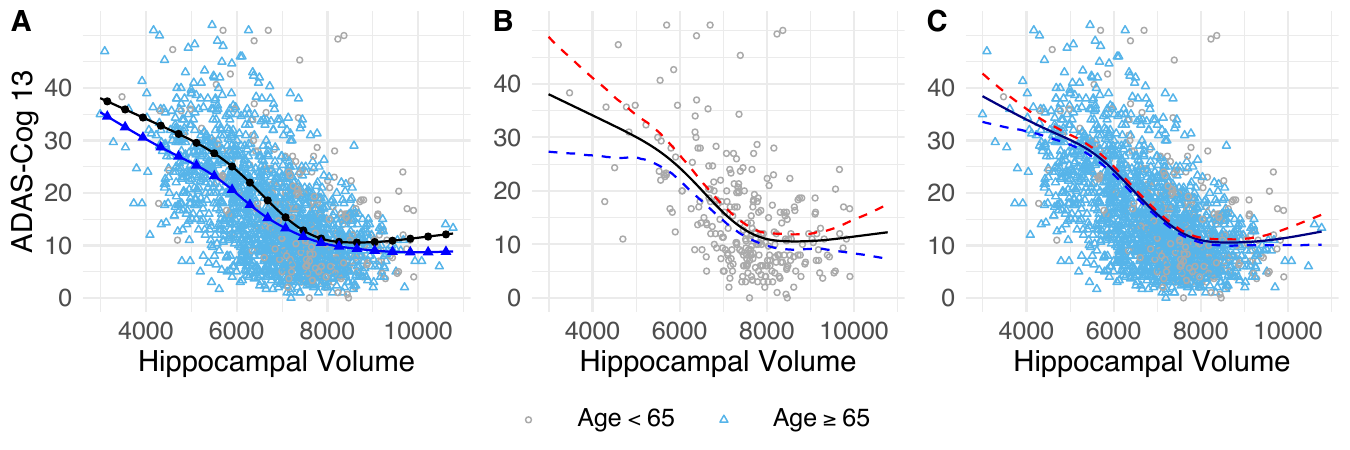}}
	\caption{(A): The solid-circle line and solid-triangle line depict the estimation of the conditional mean of ADAS-Cog 13 given the hippocampal volume based on the target data (Age$<65$) and the source data (Age$\ge65$), respectively. (B): The solid black line depicts the estimation of the conditional mean of ADAS-Cog 13 given the hippocampal volume based on target data only, and the
    dashed lines depict the 95\% confidence intervals. (C):
    The solid black line depicts the estimation of the conditional mean of ADAS-Cog 13 given the hippocampal volume based on both target and source datasets, and the
    dashed lines depict the 95\% confidence intervals.}
	\label{fig:ADNI}
\end{figure}

In our nonparametric regression model, the response variable $Y$ and the covariate $X$ are taken as ADAS-Cog 13 and hippocampal volume, respectively. ADAS-Cog 13 measures the cognitive performance (in
Alzheimer's Disease Assessment Scale-Cognitive 13-items) of an AD patient,
and hippocampal volume is measured in cm$^3$. For more information about the ADAS assessment scale, see \cite{kueper2018}.
We are more interested in the nonparametric relationship between $Y$ and $X$ for $n_0=281$ AD patients less than 65 years old. To achieve this goal, we treat such dataset as our target data, and treat the dataset consisting of $n_1=1,\!745$ AD patients over 65 years as our source data. We leverage the source data to facilitate the estimation of the target mean function of ADAS-Cog 13 given hippocampal volume for the group of AD patients less than 65 years old. Moreover, we perform statistical inference, e.g., pointwise confidence intervals, on the target mean function. 
In Figure \ref{fig:ADNI}, Panel (A) depicts the estimated target mean function and source mean function, respectively.
For each age group, there seems to be a negative relationship between 
the hippocampus volume and ADAS-Cog 13. Compared with the elder group, the effect of the hippocampus volume on ADAS-Cog 13 seems to be slightly stronger in the younger group. 
The target and source mean functions appear to be very different, which meets the considered data integration architecture.
Panel (B) depicts the estimated target mean function based on target data only, as well as the conventional 
95\% pointwise confidence intervals proposed by \cite{wahba1983jrssb}.
Panel (C) depicts the estimated target mean function based on both
target and source data, as well as the proposed 
95\% pointwise confidence intervals $\textrm{CI}^{\textrm{ds}}_{0.05}(x)$.
Clearly, the intervals in Panel (C) are much narrower than those in Panel (B) even near the boundaries (hippocampal volume approaching 4000 cm$^3$ and 10,000 cm$^3$),
demonstrating the advantage of the proposed data integration method.

\section{Conclusions}
\label{sec:conclusions}
In this paper, we investigate nonparametric inference and uncertainty quantification for the target mean function, leveraging the integration of target and source datasets. When the sizes of the target and source datasets satisfy a specific relationship, we establish the optimal rate of convergence and demonstrate bootstrap consistency for the nonparametric estimators. These results enable the construction of theoretically valid pointwise confidence intervals and confidence regions.
Our simulation studies show that the proposed method outperforms approaches relying solely on target data but underperforms when the source data overwhelmingly dominate the target data, aligning with our theoretical expectations. We apply our method to advance nonparametric inference and uncertainty quantification in Alzheimer's disease research, showcasing its practical utility.

The authors' team is currently pursuing several related projects, including extensions to advanced nonparametric domains such as functional data analysis, survival analysis, Bayesian nonparametrics, quantile regression, and isotonic regression. These projects aim to address target tasks through innovative data integration frameworks. We plan to develop statistically valid and more accurate inferential tools, and explore new real-world applications within these ongoing initiatives.

\section{Proofs}
\label{sec:proofs}
We calculate the Fr\'{e}chet derivatives of $\ell^{\textrm{cs},\omega}_{n\lambda}(f)$ as follows:
\begin{eqnarray*}
S^{\textrm{cs},\omega}_{n\lambda}(f)&\equiv&D\ell^{\textrm{cs},\omega}_{n\lambda}(f)=-\frac{1}{n}\sum_{m=0}^M\sum_{i=1}^{n_m}\omega_i^{(m)}(Y_i^{(m)}-f(X_i^{(m)}))K_{X_i^{(m)}}+\cP_\lambda f,\\
DS^{\textrm{cs},\omega}_{n\lambda}(f)\Delta f&=&\frac{1}{n}\sum_{m=0}^M\sum_{i=1}^{n_m}\omega_i^{(m)}\Delta f(X_i^{(m)})K_{X_i^{(m)}}+\cP_\lambda \Delta f,\\
D^2S^{\textrm{cs}}_{n\lambda}(f)&\equiv&0.
\end{eqnarray*}
The averaged versions are 
\begin{eqnarray*}
S^{\textrm{cs}}_\lambda(f)&\equiv&\E S^{\textrm{cs},\omega}_{n\lambda}(f)=\frac{1}{n}\sum_{m=0}^M n_m\E(f-f^{(0)}_*)(X^{(m)})K_{X^{(m)}}+\cP_\lambda f\\
&=&f-f^{(0)}_*+\cP_\lambda f^{(0)}_*,\\
DS^{\textrm{cs}}_\lambda(f)\Delta f&=&\frac{1}{n}\sum_{m=0}^M n_m \E[\Delta f(X^{(m)})K_{X^{(m)}}]+\cP_\lambda\Delta f=\Delta f,
\,\,\,\,(\textrm{hence, $DS^{\textrm{cs}}_\lambda(f)=$id})\\
D^2S^{\textrm{cs}}_\lambda(f)&=&0.
\end{eqnarray*}
It is easy to see that $DS^{\textrm{cs}}_\lambda(f)=\textrm{id}$ for any $f\in\cH$, and $S^{\textrm{cs}}_\lambda(f)=f-f^{(0)}_*+\cP_\lambda f^{(0)}_*$.
Let $f^{\textrm{cs}}_\lambda=(\textrm{id}-\cP_\lambda)f^{(0)}_*$ which satisfies $S^{\textrm{cs}}_\lambda(f^{\textrm{cs}}_\lambda)=0$.

\begin{proof}[Proof of Theorem \ref{rate:cs}]
Define $T(f)=f-S^{\textrm{cs},\omega}_{n\lambda}(f^{\textrm{cs}}_\lambda+f)$ for any $f\in\cH$.
By Taylor's expansion, we have $T(f)=-\Gamma(f)-S^{\textrm{cs},\omega}_{n\lambda}(f^{\textrm{cs}}_\lambda)$,
where 
\[
\Gamma(f)=\frac{1}{n}\sum_{m=0}^M\sum_{i=1}^{n_m}[\omega_i^{(m)}f(X_i^{(m)})K_{X_i^{(m)}}-\E f(X^{(m)})K_{X^{(m)}}].
\]
See proof of Theorem 3.1 in \cite{shang2020colt}.
The terms of $\Gamma(f)$ are almost surely bounded by $(1+c_\omega)c_1h^{-1/2}\|f\|_\infty$ in terms of $\|\cdot\|$-metric, where $c_\omega$ is an upper bound for $|\omega_i^{(m)}|$'s. 
By Theorem 3.5 of \cite{pinelis1994}, for any $f,g\in\cH$, we have
\[
\Pr\left(\|\sqrt{nh}[\Gamma(f)-\Gamma(g)]\|\ge t\right)\le 2e^{-\frac{t^2}{2(1+c_\omega)^2c_1^2\|f-g\|_\infty^2}},
\]
and so $\|\sqrt{nh}[\Gamma(f)-\Gamma(g)]\|_{\psi_2}\le 3(1+c_\omega)c_1\|f-g\|_\infty$.
Let $\|\cdot\|_{\psi_2}$ be the Orlicz norm corresponding to $\psi_2$.
Combined with Theorem 8.4 of \cite{K08} we have
\[
\bigg\|\sup_{f\in\cF}\sqrt{nh}\|\Gamma(f)\|\bigg\|_{\psi_2}\lesssim a_{\lambda}:=\int_0^1\psi_2^{-1}(D(\cF,\varepsilon,\|\cdot\|_{\infty}))d\varepsilon.
\]
By Lemma 8.1 of \cite{K08}, we have 
\begin{equation}\label{upper:sup:Gamma(f)}
\sup_{f\in\cF}\|\Gamma(f)\|=O_\Pr\left(\frac{a_\lambda}{\sqrt{nh}}\right).
\end{equation}
For any $f,g\in\cH$, it can be seen that $c_1^{-1}h^{1/2}(f-g)/\|f-g\|\in\cF$. So 
\begin{eqnarray*}
\|T(f)-T(g)\|=\|\Gamma(f-g)\|\le\sup_{f\in\cF}\|\Gamma(f)\|\times c_1h^{-1/2}\|f-g\|=O_\Pr\left(\frac{a_\lambda}{\sqrt{nh^2}}\right)\|f-g\|,    
\end{eqnarray*}
implying that $\|T(f)\|\le O_\Pr\left(\frac{a_\lambda}{\sqrt{nh^2}}\right)\|f\|+\|T(0)\|\le 2\|T(0)\|$, provided that $\|f\|\le 2\|T(0)\|=2\|S^{\textrm{cs},\omega}_{n\lambda}(f^{\textrm{cs}}_\lambda)\|$.
By contraction mapping theorem, with probability approaching one, there is $f'\in\cH$ with $\|f'\|\le 2\|S^{\textrm{cs},\omega}_{n\lambda}(f^{\textrm{cs}}_\lambda)\|$ such that $S^{\textrm{cs},\omega}_{n\lambda}(f^{\textrm{cs}}_\lambda+f')=0$, hence, $\widehat{f}^{\textrm{cs},\omega}_{n\lambda}=f^{\textrm{cs}}_\lambda+f'$.
The convergence rate follows from
\[
\|\widehat{f}^{\textrm{cs},\omega}_{n\lambda}-f^{(0)}_*\|\le \|f^{\textrm{cs}}_\lambda-f^{(0)}_*\|+\|f'\|\le\|\cP_\lambda f^{(0)}_*\|+2\|S^{\textrm{cs},\omega}_{n\lambda}(f^{\textrm{cs}}_\lambda)\|.
\]
The first term is bounded by 
\begin{equation}\label{thm:cs:rate:bias}
\|\cP_\lambda f^{(0)}_*\|=\sup_{\|f\|=1}\langle \cP_\lambda f^{(0)}_*, f\rangle=\sup_{\|f\|=1}\lambda\langle f^{(0)}_*,f\rangle_\cH\le \sqrt{\lambda}\|f^{(0)}_*\|_\cH,
\end{equation}
and the second term can be bounded by
\begin{eqnarray}
&&\E\|S^{\textrm{cs},\omega}_{n\lambda}(f^{\textrm{cs}}_\lambda)\|^2\label{thm:cs:rate:var}\\
&=&\E\|S^{\textrm{cs},\omega}_{n\lambda}(f^{\textrm{cs}}_\lambda)-S^{\textrm{cs}}_\lambda(f^{\textrm{cs}}_\lambda)\|^2\nonumber\\
&=&n^{-2}\E\|\sum_{m=0}^M\sum_{i=1}^{n_m}\omega_i^{(m)}e_i^{(m)}K_{X_i^{(m)}}\|^2\nonumber\\
&&+n^{-2}\E\|\sum_{m=0}^M\sum_{i=1}^{n_m}[\omega_i^{(m)}(f^{(0)}_*-f^{\textrm{cs}}_\lambda)(X_i^{(m)})K_{X_i^{(m)}}-\E(f^{(0)}_*-f^{\textrm{cs}}_\lambda)(X^{(m)})K_{X^{(m)}}]\|^2\nonumber\\
&\le&\frac{2c_1^2}{n^2h}\sum_{m=0}^M\sum_{i=1}^{n_m}\E|e_i^{(m)}|^2+\frac{2c_1^2}{n^2h}\sum_{m=0}^M n_m\E|f^{(0)}_*(X^{(m)})-f^{\textrm{cs}}_\lambda(X^{(m)})|^2=O\left(\frac{1}{nh}+\frac{\lambda}{nh}\right).\nonumber
\end{eqnarray}
Therefore, $\|\widehat{f}^{\textrm{cs},\omega}_{n\lambda}-f^{(0)}_*\|=O_\Pr\left(\frac{1}{\sqrt{nh}}+\sqrt{\lambda}\right)$.

Next, let $\Delta f=\widehat{f}^{\textrm{cs},\omega}_{n\lambda}-f^{(0)}_*$. By Taylor expansion,
\begin{eqnarray*}
&&S^{\textrm{cs},\omega}_{n\lambda}(\widehat{f}^{\textrm{cs},\omega}_{n\lambda})-S^{\textrm{cs},\omega}_{n\lambda}(f^{(0)}_*)-
[S^{\textrm{cs}}_{\lambda}(\widehat{f}^{\textrm{cs},\omega}_{n\lambda})-S^{\textrm{cs}}_{\lambda}(f^{(0)}_*)]\\
&=&DS^{\textrm{cs},\omega}_{n\lambda}(f^{(0)}_*)\Delta f-DS^{\textrm{cs}}_{\lambda}(f^{(0)}_*)\Delta f=\Gamma(\Delta f).
\end{eqnarray*}
Therefore,
\begin{eqnarray}
&&\|\widehat{f}^{\textrm{cs},\omega}_{n\lambda}-f^{(0)}_*+S^{\textrm{cs},\omega}_{n\lambda}(f^{(0)}_*)\|=\|\Gamma(\Delta f)\|
\le \sup_{f\in\cF}\|\Gamma(f)\|\times c_1h^{-1/2}\|\Delta f\|\\
&=&O_\Pr\left(\frac{a_\lambda}{\sqrt{nh^2}}\left(\frac{1}{\sqrt{nh}}+\sqrt{\lambda}\right)\right).\label{bhd:cs:rate}
\end{eqnarray}
This completes the proof.
\end{proof}

The proof of Theorem \ref{rate:ds} proceeds in two stages. First, we derive the convergence rate for $\widehat{f}^{(m)}$ similar to the proof of Theorem \ref{rate:cs}. Second, we derive the convergence rate for $\widehat{f}^{\textrm{ds},\omega}_{n\lambda}$ using the rates from the first stage. 
The first stage is pretty similar to the proof of Theorem \ref{rate:cs}. For readability we present full details. 
We begin with calculating the Fr\'{e}chet derivatives of $\ell^{(m)}(f)$ as follows:
\begin{eqnarray*}
S^{(m)}(f)&=&-\frac{1}{n_m}\sum_{i=1}^{n_m}(Y_i^{(m)}-f(X_i^{(m)}))K^{(m)}_{X_i^{(m)}}+\cP^{(m)}_\lambda f,\\
DS^{(m)}(f)\Delta f&=&\frac{1}{n_m}\sum_{i=1}^{n_m}\Delta f(X_i^{(m)})K^{(m)}_{X_i^{(m)}}+\cP^{(m)}_\lambda\Delta f, \\
D^2S^{(m)}(f)&=&0,
\end{eqnarray*}
with the averaged versions
\begin{eqnarray*}
S^{(m)}_{\lambda}(f)&\equiv&\E S^{(m)}(f)\\
&=&\E (f(X^{m})-f^{(m)}_*(X^{(m)}))K^{(m)}_{X^{(m)}}+\cP^{(m)}f=f-f^{(m)}_*+\cP^{(m)}_\lambda f^{(m)}_*,\\
DS^{(m)}_{\lambda}(f)\Delta f&=&\E\Delta f(X^{(m)})K^{(m)}_{X^{(m)}}+\cP^{(m)}_\lambda\Delta f=\Delta f,\,\,\,\, \textrm{(hence, $DS^{(m)}_{\lambda}(f)=$id)}\\
D^2S^{(m)}_{\lambda}(f)&=&0.
\end{eqnarray*}
Let $f^{(m)}_{\lambda}=(\textrm{id}-\cP^{(m)}_\lambda)f^{(m)}_*$ so that $S^{(m)}_{\lambda}(f^{(m)}_{\lambda})=0$.
Similar to Theorem \ref{rate:cs}, we have the following proposition.
\begin{proposition}\label{prop:ds:stage1}
Suppose that Assumptions \ref{A1}, \ref{A2}, \ref{A3} hold, and $a_\lambda=o(\sqrt{n_mh^2})$. Then we have, as $n_m\to\infty$, 
\begin{eqnarray*}
\|\widehat{f}^{(m)}-f^{(m)}_*\|_m&=&O_\Pr\left(\frac{1}{\sqrt{n_mh}}+\sqrt{\lambda}\right),\\
\|\widehat{f}^{(m)}-f^{(m)}_*+S^{(m)}(f^{(m)}_*)\|_m&=&O_\Pr\left(\frac{a_\lambda}{\sqrt{n_mh^2}}\left(\frac{1}{\sqrt{n_mh}}+\sqrt{\lambda}\right)\right).
\end{eqnarray*}
\end{proposition}
\begin{proof}[Proof of Proposition \ref{prop:ds:stage1}]
Let $T_m(f)=f-S^{(m)}(f+f^{(m)}_\lambda)$. Similar to the proof of Theorem \ref{rate:cs}, one can show that
\[
\sup_{f\in\cF}\|\Gamma_m(f)\|=O_\Pr\left(\frac{a_\lambda}{\sqrt{n_mh}}\right),
\]
where $\Gamma_m(f)=\frac{1}{n_m}\sum_{i=1}^{n_m}[f(X_i^{(m)})K_{X_i^{(m)}}-\E f(X^{(m)})K_{X^{(m)}}]$,
which leads to that $T_m$ is a contraction mapping from $\mathbb{B}(0,2\|S^{(m)}(f^{(m)}_\lambda)\|_m,\|\cdot\|_m)$ to itself, where $\mathbb{B}(0,r,\|\cdot\|_m)$ is a centered $r$-ball of $\cH$ in terms of $\|\cdot\|_m$-metric.    
\end{proof}

Next we observe that 
\begin{eqnarray*}
\widetilde{Y}_i^{(m)}&=&Y_i^{(m)}-\widehat{\delta}^{(m)}(X_i^{(m)})\\
&=&f^{(m)}_*(X_i^{(m)})+e_i^{(m)}-\widehat{f}^{(m)}(X_i^{(m)})+\widehat{f}^{(0)}(X_i^{(m)})\\
&=&f^{(0)}_*(X_i^{(m)})+r^{(0)}(X_i^{(m)})-r^{(m)}(X_i^{(m)})+e_i^{(m)}\equiv f^{(0)}_*(X_i^{(m)})+\Delta_i^{(m)},
\end{eqnarray*}
where $r^{(m)}=\widehat{f}^{(m)}-f^{(m)}_*$.
Let 
\[
\Delta_n(\omega)=-\frac{1}{n}\sum_{m=0}^M\sum_{i=n_m+1}^{2n_m}\omega_i^{(m)}\Delta_i^{(m)}K_{X_i^{(m)}}.
\]
We can calculate the Fr\'{e}chet derivatives of $\ell^{\textrm{ds},\omega}_{n\lambda}(f)$ as follows:
\begin{eqnarray*}
S^{\textrm{ds},\omega}_{n\lambda}(f)&\equiv&D\ell^{\textrm{ds},\omega}_{n\lambda}(f)\\
&=&\Delta_n(\omega)+\frac{1}{n}\sum_{m=0}^M\sum_{i=n_m+1}^{2n_m}\omega_i^{(m)}(f-f^{(0)}_*)(X_i^{(m)})K_{X_i^{(m)}}+\cP_\lambda f,\\
DS^{\textrm{ds},\omega}_{n\lambda}(f)\Delta f&=&\frac{1}{n}\sum_{m=0}^M\sum_{i=n_m+1}^{2n_m}\omega_i^{(m)}\Delta f(X_i^{(m)})K_{X_i^{(m)}}+\cP_\lambda\Delta f,\\
D^2S^{\textrm{ds},\omega}_{n\lambda}(f)&=&0,
\end{eqnarray*}
with the averaged versions (conditional on $\cD\equiv\{(Y_i^{(m)},X_i^{(m)}): i=1,\ldots,n_m, m=0,\ldots,M\}$) calculated as
\begin{eqnarray*}
S^{\textrm{ds}}_\lambda(f)&\equiv&\E[S^{\textrm{ds},\omega}_{n\lambda}(f)|\cD]
=\E[\Delta_n(\omega)|\cD]+\frac{1}{n}\sum_{m=0}^M n_m\E(f-f^{(0)}_*)(X^{(m)})K_{X^{(m)}}+\cP_\lambda f\\
&=&\E[\Delta_n(\omega)|\cD]+f-f^{(0)}_*+\cP_\lambda f^{(0)}_*,\\
DS^{\textrm{ds}}_\lambda(f)&=&\textrm{id},\\
D^2S^{\textrm{ds}}_\lambda(f)&=&0.
\end{eqnarray*}

\begin{proof}[Proof of Theorem \ref{rate:ds}]
Let $f^{\textrm{ds}}_\lambda=(\textrm{id}-\cP_\lambda)f^{(0)}_*-\E[\Delta_n(\omega)|\cD]$, then $S^{\textrm{ds}}_\lambda(f^{\textrm{ds}}_\lambda)=0$. 
Let $T(f)=f-S^{\textrm{ds},\omega}_{n\lambda}(f+f^{\textrm{ds}}_\lambda)$ for $f\in\cH$.
Similar to Theorem \ref{rate:cs}, we have $T(f)=-\Gamma(f;\omega)-S^{\textrm{ds},\omega}_{n\lambda}(f^{\textrm{ds}}_\lambda)$, where
\[
\Gamma(f;\omega)=\frac{1}{n}\sum_{m=0}^M\sum_{i=n_m+1}^{2n_m}[\omega_i^{(m)}f(X_i^{(m)})K_{X_i^{(m)}}-\E f(X^{(m)})K_{X^{(m)}}].
\]
Similar to (\ref{upper:sup:Gamma(f)}) we have 
\[
\sup_{f\in\cF}\|\Gamma(f;\omega)\|=O_\Pr\left(\frac{a_\lambda}{\sqrt{nh}}\right).
\]
So with probability approaching one, $T$ is a contraction mapping from $\mathbb{B}(0,2\|S^{\textrm{ds},\omega}_{n\lambda}(f^{\textrm{ds}}_\lambda)\|,\|\cdot\|)$ to itself. By contraction mapping theorem, 
there exists a unique $f''\in\cH$ with $\|f''\|\le 2\|S^{\textrm{ds},\omega}_{n\lambda}(f^{\textrm{ds}}_\lambda)\|$
and $T(t'')=f''$, implying that $\widehat{f}^{\textrm{ds},\omega}_{n\lambda}=f^{\textrm{ds}}_\lambda+f''$.
Therefore, 
\begin{eqnarray*}
\|\widehat{f}^{\textrm{ds},\omega}_{n\lambda}-f^{(0)}_*\|&\le&\|f^{\textrm{ds}}_\lambda-f^{(0)}_*\|+2\|S^{\textrm{ds},\omega}_{n\lambda}(f^{\textrm{ds}}_\lambda)\|\\
&\le&\|\E[\Delta_n(\omega)|\cD]\|+\|\cP_\lambda f^{(0)}_*\|+2\|S^{\textrm{ds},\omega}_{n\lambda}(f^{\textrm{ds}}_\lambda)\|.
\end{eqnarray*}
Similar to (\ref{thm:cs:rate:bias}), the second term is upper bounded by $\sqrt{\lambda}\|f^{(0)}_*\|_\cH$.
To analyze the third term, note that
\begin{eqnarray*}
\|S^{\textrm{ds},\omega}_{n\lambda}(f^{\textrm{ds}}_\lambda)\|&=&\|S^{\textrm{ds},\omega}_{n\lambda}(f^{\textrm{ds}}_\lambda)-S^{\textrm{ds}}_\lambda(f^{\textrm{ds}}_\lambda)\|\\
&\le&\|\Delta_n(\omega)-\E[\Delta_n(\omega)|\cD]\|\\
&&+\|\frac{1}{n}\sum_{m=0}^M\sum_{i=n_m+1}^{2n_m}[\omega_i^{(m)}(f^{\textrm{ds}}_\lambda-f^{(0)}_*)(X_i^{(m)})-\E_{X^{(m)}}(f^{\textrm{ds}}_\lambda-f^{(0)}_*)(X^{(m)})K_{X^{(m)}}]\|.
\end{eqnarray*}
Similar to (\ref{thm:cs:rate:var}), the conditional expectation of the square of the second term given $\cD$ is upper bounded by
\begin{eqnarray*}
&&\frac{1}{n^2}\sum_{m=0}^M \sum_{i=n_m+1}^{2n_m}\E[\|\omega_i^{(m)}(f^{\textrm{ds}}_\lambda-f^{(0)}_*)(X_i^{(m)})K_{X_i^{(m)}}\|^2|\cD]\\
&=&\frac{1}{n}V(f^{\textrm{ds}}_\lambda-f^{(0)}_*)
\le\frac{1}{n}\|f^{\textrm{ds}}_\lambda-f^{(0)}_*\|^2\le \frac{2}{n}\|\E[\Delta_n(\omega)|\cD]\|^2+\frac{2\lambda}{n}\|f^{(0)}_*\|_\cH^2.
\end{eqnarray*}
Therefore,
\[
\|S^{\textrm{ds},\omega}_{n\lambda}(f^{\textrm{ds}}_\lambda)\|\le \|\Delta_n(\omega)-\E[\Delta_n(\omega)|\cD]\|+O_\Pr\left(\frac{\|\E[\Delta_n(\omega)|\cD]\|}{\sqrt{n}}+\sqrt{\frac{\lambda}{n}}\right).
\]
So it is sufficient to bound $\|\Delta_n(\omega)-\E[\Delta_n(\omega)|\cD]\|$ and $\|\E[\Delta_n(\omega)|\cD]\|$.
Observe that
\begin{eqnarray*}
&&\|\Delta_n(\omega)-\E[\Delta_n(\omega)|\cD]\|\\
&\le&\frac{1}{n}\|\sum_{m=0}^M\sum_{i=n_m+1}^{2n_m}\omega_i^{(m)}e_i^{(m)}K_{X_i^{(m)}}\|\\
&&+\frac{1}{n}\|\sum_{m=1}^M\sum_{i=n_m+1}^{2n_m}\left(\omega_i^{(m)}(r^{(0)}-r^{(m)})(X_i^{(m)})K_{X_i^{(m)}}-\E[(r^{(0)}-r^{(m)})(X^{(m)})K_{X^{(m)}}|\cD]\right)\|.
\end{eqnarray*}
By Assumption \ref{A3}, the square of the first term has expectation bounded by $O((nh)^{-1})$.
The square of the second term has conditional expectation given $\cD$ bounded by
\[
\frac{c_1^2}{n^2h}\sum_{m=1}^M n_m V_m(r^{(0)}-r^{(m)})\le \frac{c_1^2}{n^2h}\sum_{m=1}^M n_m\|r^{(0)}-r^{(m)}\|_m^2.
\]
By Proposition \ref{prop:ds:stage1} and (\ref{norm:equiv:cs}), and rate condition $\frac{1}{\sqrt{n_mh}}+\sqrt{\lambda}=O(1)$, the RHS is further bounded by
\[
\frac{2bc_1^2}{nh}\|r^{(0)}\|_0^2+\frac{2c_1^2}{n^2h}\sum_{m=1}^M n_m\|r^{(m)}\|_m^2=O_\Pr((nh)^{-1}).
\]
So we have $\|\Delta_n(\omega)-\E[\Delta_n(\omega)|\cD]\|=O_\Pr((nh)^{-1/2})$.

Now we analyze $\|\E[\Delta_n(\omega)|\cD]\|$. Let 
\[
\eta^{(m)}=\widehat{f}^{(m)}-f^{(m)}_*+S^{(m)}(f^{(m)}_*),\,\,m=0,\ldots,M.
\]
Then $r^{(m)}=\eta^{(m)}-S^{(m)}(f^{(m)}_*)$, and by Proposition \ref{prop:ds:stage1} and (\ref{norm:equiv:cs}), $\eta^{(m)}$ satisfies
\[
\|\eta^{(m)}\|=O_\Pr\left(\frac{a_\lambda}{\sqrt{n_mh^2}}\left(\frac{1}{\sqrt{n_mh}}+\sqrt{\lambda}\right)\right).
\]
So we have
\begin{eqnarray*}
\E[\Delta_n(\omega)|\cD]&=&-\frac{1}{n}\sum_{m=1}^M n_m\E[(r^{(0)}-r^{(m)})(X^{(m)})K_{X^{(m)}}|\cD]\\
&=&-\frac{1}{n}\sum_{m=1}^M n_m\E[(\eta^{(0)}-\eta^{(m)})(X^{(m)})K_{X^{(m)}}|\cD]\\
&&-\frac{1}{n}\sum_{m=1}^M n_m \E[(S^{(m)}(f^{(m)}_*)-S^{(0)}(f^{(0)}_*))(X^{(m)})K_{X^{(m)}}|\cD]\\
&\equiv&
-\frac{1}{n}\sum_{m=1}^M n_m \delta_{m1}-\frac{1}{n}\sum_{m=1}^M n_m \delta_{m2}.
\end{eqnarray*}
Observe that
\begin{eqnarray*}
\|\delta_{m1}\|&\le&\|\E[(\eta^{(0)}-\eta^{(m)})(X^{(m)})K_{X^{(m)}}|\cD]\|\\
&\le&\E[|(\eta^{(0)}-\eta^{(m)})(X^{(m)})|\cdot\|K_{X^{(m)}}\||\cD]\\
&\le&c_1h^{-1/2}\E[|(\eta^{(0)}-\eta^{(m)})(X^{(m)})|\,\,|\cD]\\
&\le&c_1h^{-1/2}[\sqrt{b}\|\eta^{(0)}\|_0+\|\eta^{(m)}\|_m]\\
&=&c_1h^{-1/2}\left[O_\Pr\left(\frac{a_\lambda}{\sqrt{n_0h^2}}\left(\frac{1}{\sqrt{n_0h}}+\sqrt{\lambda}\right)\right)+
O_\Pr\left(\frac{a_\lambda}{\sqrt{n_mh^2}}\left(\frac{1}{\sqrt{n_mh}}+\sqrt{\lambda}\right)\right)\right]\\
&=&O_\Pr((nh)^{-1/2}),
\end{eqnarray*}
where the last equality follows from the rate condition $\frac{a_\lambda}{\sqrt{n_mh^2}}\left(\frac{1}{\sqrt{n_mh}}+\sqrt{\lambda}\right)=O(n^{-1/2})$, leading to $\|\frac{1}{n}\sum_{m=1}^M n_m \delta_{m1}\|=O_\Pr((nh)^{-1/2})$.
Meanwhile, 
\begin{eqnarray*}
\frac{1}{n}\sum_{m=1}^M n_m\delta_{m2}&=&\frac{1}{n}\sum_{m=1}^M n_m \E[S^{(m)}(f^{(m)}_*)(X^{(m)})K_{X^{(m)}}|\cD]\\
&&-\frac{1}{n}\sum_{m=1}^M n_m \E[S^{(0)}(f^{(0)}_*)(X^{(m)})K_{X^{(m)}}|\cD].
\end{eqnarray*}
The first term equals
\begin{eqnarray*}
-\frac{1}{n}\sum_{m=1}^M\sum_{i=1}^{n_m}e_i^{(m)}\E[K^{(m)}(X_i^{(m)},X^{(m)})K_{X^{(m)}}|\cD]+\frac{1}{n}\sum_{m=1}^M n_m\E[(\cP^{(m)}_\lambda f^{(m)}_*)(X^{(m)})K_{X^{(m)}}|\cD].    
\end{eqnarray*}
Note that
\begin{eqnarray*}
\|\E[K^{(m)}(X_i^{(m)},X^{(m)})K_{X^{(m)}}|\cD]\|&=&\sup_{\|f\|=1}\E[K^{(m)}(X_i^{(m)},X^{(m)})f(X^{(m)})|\cD]\\
&=&\sup_{\|f\|=1}V_m(K^{(m)}_{X_i^{(m)}},f)\le\sup_{\|f\|=1}\sqrt{V_m(K^{(m)}_{X_i^{(m)}})}\times\sqrt{V_m(f)}\\
&\le&b\|K^{(m)}_{X_i^{(m)}}\|_m\le bc_1h^{-1/2},
\end{eqnarray*}
and 
\begin{eqnarray*}
\|\E(\cP^{(m)}_\lambda f^{(m)}_*)(X^{(m)})K_{X^{(m)}}\|&=&\sup_{\|f\|=1}\E(\cP^{(m)}_\lambda f^{(m)}_*)(X^{(m)})f(X^{(m)})\\
&=&\sup_{\|f\|=1}V_m(\cP^{(m)}_\lambda f^{(m)}_*,f)\le \sup_{\|f\|=1}\sqrt{V_m(\cP^{(m)}_\lambda f^{(m)}_*)}\times\sqrt{V_m(f)}\\
&\le&b\|\cP^{(m)}_\lambda f^{(m)}_*\|_m\le b\sqrt{\lambda}\|f^{(m)}_*\|_\cH.
\end{eqnarray*}
So we have
\begin{eqnarray*}
&&\frac{1}{n^2}\E\|\sum_{m=1}^M\sum_{i=1}^{n_m}e_i^{(m)}\E[K^{(m)}(X_i^{(m)},X^{(m)})K_{X^{(m)}}|\cD]\|^2\\
&=&
\frac{1}{n^2}\sum_{m=1}^M\sum_{i=1}^{n_m}\E|e_i^{(m)}|^2\|\E[K^{(m)}(X_i^{(m)},X^{(m)})K_{X^{(m)}}|\cD]\|^2=O((nh)^{-1}).
\end{eqnarray*}
and
\begin{eqnarray*}
\|\frac{1}{n}\sum_{m=1}^M n_m\E[(\cP^{(m)}_\lambda f^{(m)}_*)(X^{(m)})K_{X^{(m)}}|\cD]\|=O(\sqrt{\lambda}).    
\end{eqnarray*}
The second term can be treated similarly. Hence,
$\|\frac{1}{n}\sum_{m=1}^M n_m \delta_{m2}\|=O_\Pr((nh)^{-1/2}+\sqrt{\lambda})$.
Combining the above, we have $\|\E[\Delta_n(\omega)|\cD]\|=O_\Pr((nh)^{-1/2}+\sqrt{\lambda})$.
In summary, we thus show that
$\|\widehat{f}^{\textrm{ds},\omega}_{n\lambda}-f^{(0)}_*\|=O_\Pr((nh)^{-1/2}+\sqrt{\lambda})$.
Finally, similar to (\ref{bhd:cs:rate}), it can be shown that 
\[
\|\widehat{f}^{\textrm{ds},\omega}_{n\lambda}-f^{(0)}_*+S^{\textrm{ds},\omega}_{n\lambda}(f^{(0)}_*)\|=O_\Pr\left(\frac{a_\lambda}{\sqrt{nh^2}}\left(\frac{1}{\sqrt{nh}}+\sqrt{\lambda}\right)\right).
\]
This completes the proof.
\end{proof}

\begin{proof}[Proof of Theorem \ref{thm:local:boot}]
It follows from Theorems \ref{rate:cs} and \ref{rate:ds} that 
\begin{eqnarray}
\|\widehat{f}^{\textrm{cs},\omega}_{n\lambda}-\widehat{f}^{\textrm{cs}}_{n\lambda}+S^{\textrm{cs},\omega}_{n\lambda}(f^{(0)}_*)-S^{\textrm{cs}}_{n\lambda}(f^{(0)}_*)\|
&=&O_\Pr\left(\frac{a_\lambda}{\sqrt{nh^2}}\left(\frac{1}{\sqrt{nh}}+\sqrt{\lambda}\right)\right),\label{proof:local:boot:eqn1}\\
\|\widehat{f}^{\textrm{ds},\omega}_{n\lambda}-\widehat{f}^{\textrm{ds}}_{n\lambda}+S^{\textrm{ds},\omega}_{n\lambda}(f^{(0)}_*)-S^{\textrm{ds}}_{n\lambda}(f^{(0)}_*)\|
&=&O_\Pr\left(\frac{a_\lambda}{\sqrt{nh^2}}\left(\frac{1}{\sqrt{nh}}+\sqrt{\lambda}\right)\right).\label{proof:local:boot:eqn2}
\end{eqnarray}
We only prove Part (1) since Part (2) can be done similarly. 
We fix an $x\in\cX$. By CLT, given the data $(Y_i^{(m)},X_i^{(m)})$'s, as $n\to\infty$,
\begin{eqnarray*}
\frac{1}{\tau_n(x)}\left[S^{\textrm{cs},\omega}_{n\lambda}(f^{(0)}_*)(x)-S^{\textrm{cs}}_{n\lambda}(f^{(0)}_*)(x)\right]
&=&-\frac{1}{n\tau_n(x)}\sum_{m=0}^M\sum_{i=1}^{n_m}(\omega_i^{(m)}-1)e_i^{(m)}K(X_i^{(m)},x)\\
&\overset{\bD}{\to}&N(0,1),
\end{eqnarray*}
where $\tau_n^2(x):=\frac{1}{n^2}\sum_{m=0}^M\sum_{i=1}^{n_m}|e_i^{(m)}|^2|K(X_i^{(m)},x)|^2=O_\Pr((nh)^{-1})$.
Therefore, by (\ref{bound:K_x}) and rate condition $\frac{a_\lambda}{h}\left(\frac{1}{\sqrt{nh}}+\sqrt{\lambda}\right)=o(1)$, as $n\to\infty$,
\begin{equation}\label{local:clt:1}
\frac{1}{\tau_n(x)}\left[\widehat{f}^{\textrm{cs},\omega}_{n\lambda}(x)-\widehat{f}^{\textrm{cs}}_{n\lambda}(x)\right]\overset{\bD}{\to}N(0,1).
\end{equation}
Let $\tau^2(x):=\E\tau_n^2(x)$ which has an order $O((nh)^{-1})$.
By CLT and the rate condition  
$\sqrt{nh}(\cP_\lambda f^{(0)}_*)(x)=o(1)$ (which implies $(\cP_\lambda f^{(0)}_*)(x)/\tau(x)=(1)$), we have as $n\to\infty$,
\begin{eqnarray}
&&\frac{1}{\tau(x)}\left[\widehat{f}^{\textrm{cs}}_{n\lambda}(x)-f^{(0)}_*(x)\right]\nonumber\\
&=&\frac{1}{n\tau(x)}\sum_{m=0}^M\sum_{i=1}^{n_m}e_i^{(m)}K(X_i^{(m)},x)-\frac{(\cP_\lambda f^{(0)}_*)(x)}{\tau(x)}+O_\Pr\left(\frac{a_\lambda}{h}\left(\frac{1}{\sqrt{nh}}+\sqrt{\lambda}\right)\right)\nonumber\\
&\overset{\bD}{\to}&N(0,1).\label{local:clt:2}
\end{eqnarray}
Combining (\ref{local:clt:1}) and (\ref{local:clt:2}), for any $u\in\bbR$ we have
\begin{eqnarray*}
\Pr_*\left(\widehat{f}^{\textrm{cs},\omega}_{n\lambda}(x)-\widehat{f}^{\textrm{cs}}_{n\lambda}(x)\le\tau_n(x)u\right)&\overset{\Pr}{\to}&\Phi(u),\\    
\Pr\left(\widehat{f}^{\textrm{cs}}_{n\lambda}(x)-f^{(0)}_*(x)\le \tau(x)u\right)&\to&\Phi(u).
\end{eqnarray*}
Proof is complete.
\end{proof}

\begin{proof}[Proof of Theorem \ref{thm:global:boot}]
We only prove Part (1) since Part (2) can be done similarly. 
We still begin with (\ref{proof:local:boot:eqn1}) and (\ref{proof:local:boot:eqn2}).
By (\ref{proof:local:boot:eqn1}) and triangle inequality, we have
\begin{eqnarray*}
\sqrt{V(\widehat{f}^{\textrm{cs},\omega}_{n\lambda}-\widehat{f}^{\textrm{cs}}_{n\lambda})}&=&\sqrt{V(S^{\textrm{cs},\omega}_{n\lambda}(f^{(0)}_*)-S^{\textrm{cs}}_{n\lambda}(f^{(0)}_*))}+O_\Pr\left(\frac{a_\lambda}{\sqrt{nh^2}}\left(\frac{1}{\sqrt{nh}}+\sqrt{\lambda}\right)\right).
\end{eqnarray*}
By direct calculations,
\begin{eqnarray}
&&V(S^{\textrm{cs},\omega}_{n\lambda}(f^{(0)}_*)-S^{\textrm{cs}}_{n\lambda}(f^{(0)}_*))\\
&=&\frac{1}{n^2}\sum_{m,s=0}^M\sum_{i=1}^{n_m}\sum_{j=1}^{n_s}(\omega_i^{(m)}-1)(\omega_j^{(s)}-1)e_i^{(m)}e_j^{(s)}
V(K_{X_i^{(m)}},K_{X_j^{(s)}})\nonumber\\
&\equiv&\frac{1}{n^2}\sum_{m,s=0}^M\sum_{i=1}^{n_m}\sum_{j=1}^{n_s}W_{m,i;s,j}\nonumber\\
&=&\frac{1}{n^2}\sum_{m=0}^M\sum_{i=1}^{n_m}W_{m,i;m,i}+\frac{2}{n^2}\sum_{(m,i;s,j)\in\cN}W_{m,i;s,j}\equiv W_1+W_2,\label{proof:glb:boot:eqn1}
\end{eqnarray}
where $\cN$ is a subset of $\{(m,i;s,j): m,s=0,\ldots,M, i,j=1,\ldots,n_m, (m,i)\neq (s,j)\}$ such that, if $(m,i;s,j)\in\cN$ then $(s,j;m,i)\notin\cN$.
The conditional expectation of $W_1$ given $\cD$ equals 
\[
\sigma_n^2:=\E[W_1|\cD]=\frac{1}{n^2}\sum_{m=0}^M\sum_{i=1}^{n_m}|e_i^{(m)}|^2V(K_{X_i^{(m)}},K_{X_i^{(m)}}).
\]
Since $\omega^{(m)}$ has finite fourth order moments,
\begin{eqnarray*}
\E[|W_1-\E[W_1|\cD]|^2|\cD]&\le&\frac{1}{n^4}\sum_{m=0}^M\sum_{i=1}^{n_m}\E[W_{(m,i;m,i)}^2|\cD]\\
&\lesssim& 
\frac{1}{n^4}\sum_{m=0}^M\sum_{i=1}^{n_m} |e_i^{(m)}|^4V(K_{X_i^{(m)}},K_{X_i^{(m)}})^2\\
&\le&\frac{\max_{m,i}|e_i^{(m)}|^2}{n^2h}\sigma_n^2:=\frac{U_n}{n^2h}\sigma_n^2.
\end{eqnarray*}
So $W_1-\E[W_1|\cD]=O_{\Pr_*}\left(\sqrt{\frac{U_n}{n^2h}\sigma_n^2}\right)$.
By the condition that $\E[\exp\left(\kappa|e^{(m)}|^2\right)]<\infty$ for some constant $\kappa>1$, we have $U_n=O_\Pr(\log{n})$. 
Let $\sigma^2:=\E\sigma_n^2$. Note that 
\begin{equation}\label{eqn:sigma:sq}
\sigma^2=\frac{1}{n^2}\sum_{m=0}^M n_m\E[|e^{(m)}|^2 V(K_{X^{(m)}},K_{X^{(m)}})]\asymp\frac{1}{nh}.
\end{equation}
By direct calculations, 
\[
\E[|\sigma_n^2-\sigma^2|^2]\le\frac{1}{n^4}\sum_{m=0}^M n_m\E[|e^{(m)}|^4V(K_{X^{(m)}},K_{X^{(m)}})^2]=O(n^{-3}h^{-2}),
\]
we have $\sigma_n^2=(1+o_\Pr(1))\sigma^2$, and so $\sigma_n^2=O_\Pr((nh)^{-1})$. 
Meanwhile, the conditional variance of $W_2$ given $\cD$ is 
\[
\zeta_n^2:=\E[W_2^2|\cD]=\frac{4}{n^4}\sum_{(m,i;s,j)\in\cN}|e_i^{(m)}|^2|e_j^{(s)}|^2V(K_{X_i^{(m)}},K_{X_j^{(s)}})^2\equiv \frac{4}{n^4}\sum_{(m,i;s,j)\in\cN}Z_{(m,i;s,j)}.
\]
This holds since the cross terms all vanish. That is, if $(m,i;s,j)\neq(m',i';s',j')$, then 
\[
\E[W_{(m,i;s,j)}W_{(m',i';s',j')}|\cD]=0.
\]
One can check this in three cases: (1) $(m,i)=(m',i')$ and $(s,j)\neq(s',j')$;
(2) $(m,i)\neq(m',i')$ and $(s,j)=(s',j')$; (3) $(m,i)\neq(m',i')$ and $(s,j)\neq(s',j')$.
Cases (1) and (2) are straightforward thanks to independence. Note that (3) excludes the possibility that $(m,i)=(s',j')$ and $(m',i')=(s,j)$; otherwise both $(m,i;s,j)$ and $(s,j;m,i)$ would belong to $\cN$ which is impossible by definition of $\cN$.

We further analyze how $\zeta_n^2$ is concentrated around its mean $\zeta^2:=\E\zeta_n^2$.
Note that
\begin{eqnarray*}
&&\E[|\zeta_n^2-\E\zeta_n^2|^2]\\
&=&\frac{16}{n^8}\sum_{(m,i;s,j),(m',i';s',j')\in\cN}\left(\E[Z_{(m,i;s,j)}Z_{(m',i';s',j')}]-\E[Z_{(m,i;s,j)}]\E[Z_{(m',i';s',j')}]\right).
\end{eqnarray*}
For $(m,i;s,j),(m',i';s',j')\in\cN$ with pairwise different $(m,i),(s,j),(m',i'),(s',j')$, we have 
\[
\E[Z_{(m,i;s,j)}Z_{(m',i';s',j')}]-\E[Z_{(m,i;s,j)}]\E[Z_{(m',i';s',j')}]=0
\]
due to independence.
Hence, we only consider the case that some pairs are the same.
There are at most $O(n^3)$ such terms with each bounded by $O(h^{-4})$;
by Assumption 3, the eighth moments of $e^{(m)}$'s are finitely
bounded so this holds.
So $\zeta_n^2-\E\zeta_n^2=O_\Pr(n^{-5/2}h^{-2})$.
On the other hand,
\begin{eqnarray*}
\E\zeta_n^2&\asymp&\frac{1}{n^4}\sum_{(m,i;s,j)\in\cN}
\E[|e_i^{(m)}|^2|e_j^{(s)}|^2V(K_{X_i^{(m)}},K_{X_j^{(s)}})^2]\\
&\asymp&\frac{1}{n^4}\sum_{(m,i;s,j)\in\cN}\E[V(K_{X_i^{(m)}},K_{X_j^{(s)}})^2].
\end{eqnarray*}
Let $(\phi_\nu,\theta_\nu)$'s be eigen-pairs such that $V(\phi_\nu,\phi_\mu)=\delta_{\nu\mu}$,
$\langle\phi_\nu,\phi_\mu\rangle_\cH=\theta_\nu\delta_{\nu\mu}$, and $K$ has an expression $K(x,x')=\sum_\nu\frac{\phi_\nu(x)\phi_\nu(x')}{1+\lambda\theta_\nu}$.
Then 
\begin{eqnarray*}
&&\frac{1}{n^4}\sum_{(m,i;s,j)\in\cN}
\E[V(K_{X_i^{(m)}},K_{X_j^{(s)}})^2]\\
&=&\frac{1}{n^4}\sum_{(m,i;s,j)\in\cN}\sum_{\nu,\mu}\frac{\E\phi_\nu(X_i^{(m)})\phi_\mu(X_i^{(m)})
\phi_\nu(X_j^{(s)})\phi_\mu(X_j^{(s)})}{(1+\lambda\theta_\nu)^2(1+\lambda\theta_\mu)^2}\\
&=&\frac{1}{n^4}\sum_{(m,i;s,j)\in\cN}\sum_{\nu,\mu}\frac{V_m(\phi_\nu,\phi_\mu)V_s(\phi_\nu,\phi_\mu)}{(1+\lambda\theta_\nu)^2(1+\lambda\theta_\mu)^2}\\
&\asymp&\frac{1}{n^2}\sum_{\nu,\mu}\frac{V(\phi_\nu,\phi_\mu)^2}{(1+\lambda\theta_\nu)^2(1+\lambda\theta_\mu)^2}\asymp n^{-2}h^{-1},
\end{eqnarray*}
where the last step follows from Assumption \ref{A1} which guarantees the equivalence of $V_0$ and $V$, as well as mapping principle in Section 3 of \cite{weinberg1974} which says that $\theta_\nu$ and $\rho_\nu$ are equivalent. Due to the rate condition $nh^2\gg1$, we thus have $\zeta_n^2=(1+o_\Pr(1))\E\zeta_n^2$.

It can be examined that 
\begin{eqnarray*}
G_I&:=&\frac{1}{n^8}\sum_{(m,i;s,j)\in\cN}\E[W_{(m,i;s,j)}^4|\cD]\\
&\lesssim&\frac{1}{n^8}\sum_{(m,i;s,j)\in\cN}|e_i^{(m)}|^4|e_j^{(s)}|^4V(K_i^{(m)},K_j^{(s)})^4\lesssim\frac{U_n^2}{n^4h^2}\zeta_n^2.
\end{eqnarray*}
\begin{eqnarray*}
G_{II}&:=&\frac{1}{n^8}\sum_{\textrm{$(m,i),(s,j),(t,k)$ are p.d.}}\left\{\E[W_{(m,i;s,j)}^2W_{(m,i;t,k)}^2|\cD]\right.\\
&&\left.+\E[W_{(s,j;m,i)}^2W_{(s,j;t,k)}^2|\cD]+\E[W_{(t,k;m,i)}^2W_{(t,k;s,j)}^2|\cD]\right\},
\end{eqnarray*}
where ``p.d.'' represents ``pairwise different.''
Since 
\begin{eqnarray*}
&&\sum_{\textrm{$(m,i),(s,j),(t,k)$ are p.d.}} \E[W_{(m,i;s,j)}^2W_{(m,i;t,k)}^2|\cD]\\
&\le&\sum_{\textrm{$(m,i),(s,j),(t,k)$ are p.d.}} \sqrt{\E[W_{(m,i;s,j)}^4|\cD]}\sqrt{\E[W_{(m,i;t,k)}^4|\cD]} \\
&\lesssim&\sum_{\textrm{$(m,i),(s,j),(t,k)$ are p.d.}}|e_i^{(m)}|^4|e_j^{(s)}|^2|e_k^{(t)}|^2V(K_{X_i^{(m)}},K_{X_j^{(s)}})^2V(K_{X_i^{(m)}},K_{X_k^{(t)}})^2\\
&\lesssim& U_n^2h^{-2}\sum_{\textrm{$(m,i),(s,j),(t,k)$ are p.d.}}|e_i^{(m)}|^2|e_j^{(s)}|^2V(K_{X_i^{(m)}},K_{X_j^{(s)}})^2\\
&\le&U_n^2nh^{-2}\sum_{(m,i)\neq (s,j)}|e_i^{(m)}|^2|e_j^{(s)}|^2V(K_{X_i^{(m)}},K_{X_j^{(s)}})^2=\frac{1}{2}U_n^2n^5h^{-2}\zeta_n^2.
\end{eqnarray*}
Other terms in $G_{II}$ could be bounded similarly. Therefore, $G_{II}\lesssim \frac{U_n^2}{n^3h^2}\zeta_n^2$.
Due to the rate condition $nh\gg(\log{n})^2$, we have both $G_I=O_\Pr(\zeta_n^4)$ and $G_{II}=O_\Pr(\zeta_n^4)$.
Next, we have
\begin{eqnarray*}
&&G_{IV}:=\\
&&\frac{1}{n^8}\sum_{\textrm{$(m,i),(s,j),(t,k), (r,l)$ are p.d.}}\left\{\E[W_{(m,i;s,j)}W_{(m,i;t,k)}W_{(r,l;s,j)}W_{(r,l;t,k)}|\cD]\right.\\
&&\left.+\E[W_{(m,i;s,j)}W_{(m,i;r,l)}W_{(t,k;s,j)}W_{(t,k;r,l)}|\cD]+\E[W_{(m,i;t,k)}W_{(m,i;r,l)}W_{(s,j;t,k)}W_{(s,j;r,l)}|\cD]\right\}.    
\end{eqnarray*}
For pairwise different $(m,i),(s,j),(t,k), (r,l)$, we have
\begin{eqnarray*}
&&\E[W_{(m,i;s,j)}W_{(m,i;t,k)}W_{(r,l;s,j)}W_{(r,l;t,k)}]\\
&\lesssim&
\E[V(K_{X_i^{(m)}},K_{X_j^{(s)}})V(K_{X_i^{(m)}},K_{X_k^{(t)}})V(K_{X_l^{(r)}},K_{X_j^{(s)}})V(K_{X_l^{(r)}},K_{X_k^{(t)}})]\\
&=&\sum_{\mu_1,\mu_2,\mu_3,\mu_4}\frac{V_m(\phi_{\mu_1},\phi_{\mu_2})V_s(\phi_{\mu_1},\phi_{\mu_3})V_t(\phi_{\mu_2},\phi_{\mu_4})V_r(\phi_{\mu_3},\phi_{\mu_4})}{(1+\lambda\theta_{\mu_1})^2(1+\lambda\theta_{\mu_2})^2(1+\lambda\theta_{\mu_3})^2(1+\lambda\theta_{\mu_4})^2},
\end{eqnarray*}
so we have
\begin{eqnarray*}
\E[G_{IV}]&\lesssim&n^{-4}\sum_{\mu_1,\mu_2,\mu_3,\mu_4}\frac{V(\phi_{\mu_1},\phi_{\mu_2})V(\phi_{\mu_1},\phi_{\mu_3})
V(\phi_{\mu_2},\phi_{\mu_4})V(\phi_{\mu_3},\phi_{\mu_4})}{(1+\lambda\theta_{\mu_1})^2(1+\lambda\theta_{\mu_2})^2(1+\lambda\theta_{\mu_3})^2(1+\lambda\theta_{\mu_4})^2}\asymp \frac{1}{n^4h},    
\end{eqnarray*}
implying that $G_{IV}=O_\Pr(n^{-4}h^{-1})=o_\Pr(\zeta_n^4)$.
By a conditional version of Proposition 3.2 of \cite{de1987central},
we have $W_2/\zeta_n\overset{\bD}{\to}N(0,1)$ in $\Pr_*$-probability.
Recall that $W_1-\sigma_n^2=O_{\Pr_*}\left(\sqrt{\frac{\log{n}}{n^3h^2}}\right)=o_{\Pr_*}(\zeta_n)$,
so we have in $\Pr_*$-probability,
\[
\frac{V(S^{\textrm{cs},\omega}_{n\lambda}(f^{(0)}_*)-S^{\textrm{cs}}_{n\lambda}(f^{(0)}_*))-\sigma_n^2}{\zeta_n}
=\frac{W_1-\sigma_n^2+W_2}{\zeta_n}=o_{\Pr_*}(1)+\frac{W_2}{\zeta_n}\overset{\bD}{\to}N(0,1).
\]
This also implies $V(S^{\textrm{cs},\omega}_{n\lambda}(f^{(0)}_*)-S^{\textrm{cs}}_{n\lambda}(f^{(0)}_*)=O_\Pr((nh)^{-1})$.
By rate condition $\frac{a_\lambda}{h}\left(\frac{1}{\sqrt{nh}}+\sqrt{\lambda}\right)=o(1)$ we get
\begin{eqnarray*}
V(\widehat{f}^{\textrm{cs},\omega}_{n\lambda}-\widehat{f}^{\textrm{cs}}_{n\lambda})&=&V(S^{\textrm{cs},\omega}_{n\lambda}(f^{(0)}_*)-S^{\textrm{cs}}_{n\lambda}(f^{(0)}_*))\\
&&+O_\Pr\left(\frac{a_\lambda^2}{nh^2}\left(\frac{1}{\sqrt{nh}}+\sqrt{\lambda}\right)^2+\frac{a_\lambda}{nh\sqrt{h}}\left(\frac{1}{\sqrt{nh}}+\sqrt{\lambda}\right)\right)\\
&=&V(S^{\textrm{cs},\omega}_{n\lambda}(f^{(0)}_*)-S^{\textrm{cs}}_{n\lambda}(f^{(0)}_*))+o_\Pr(\zeta_n),
\end{eqnarray*}
which shows that for any $u\in\bbR$,
\[
\Pr_*\left(V(\widehat{f}^{\textrm{cs},\omega}_{n\lambda}-\widehat{f}^{\textrm{cs}}_{n\lambda})\le\sigma_n^2+u\zeta_n\right)
\overset{\Pr}{\to}\Phi(u).
\]

We can also treat $V(\widehat{f}^{\textrm{cs},\omega}_{n\lambda}-f^{(0)}_*)$ similarly. 
By Theorem \ref{rate:cs} and triangle inequality,
\[
\sqrt{V(\widehat{f}^{\textrm{cs}}_{n\lambda}-f^{(0)}_*)}=\sqrt{S^{\textrm{cs}}_{n\lambda}(f^{(0)}_*)}+O_\Pr\left(\frac{a_\lambda}{\sqrt{nh^2}}\left(\frac{1}{\sqrt{nh}}+\sqrt{\lambda}\right)\right).
\]
Recall that 
\[
S^{\textrm{cs}}_{n\lambda}(f^{(0)}_*)=-\frac{1}{n}\sum_{m=0}^M\sum_{i=1}^{n_m}e_i^{(m)}K_{X_i^{(m)}}+\cP_\lambda f^{(0)}_*.
\]
It can be checked that
\begin{eqnarray*}
W_3&:=&V(\frac{1}{n}\sum_{m=0}^M\sum_{i=1}^{n_m}e_i^{(m)}K_{X_i^{(m)}})=\frac{1}{n^2}\sum_{m,s=0}^M\sum_{i=1}^{n_m}\sum_{j=1}^{n_s}
e_i^{(m)}e_j^{(s)}V(K_{X_i^{(m)}},K_{X_j^{(s)}})\\
&=&\frac{1}{n^2}\sum_{m=0}^M\sum_{i=1}^{n_m}|e_i^{(m)}|^2V(K_{X_i^{(m)}},K_{X_i^{(m)}})+
\frac{2}{n^2}\sum_{(m,i;s,j)\in\cN}e_i^{(m)}e_j^{(s)}V(K_{X_i^{(m)}},K_{X_j^{(s)}})\\
&\equiv&\sigma_n^2+W_4
\end{eqnarray*}
satisfies $W_3=\sigma^2+W_4+o_\Pr(\zeta)$;
recalling that $\sigma^2$ is given in (\ref{eqn:sigma:sq}) and $\zeta^2=\E\zeta_n^2\asymp n^{-2}h^{-1}$.
By Proposition 3.2 of \cite{de1987central}, it can be shown that $W_4/\zeta\overset{\bD}{\to}N(0,1)$, hence,
$(W_3-\sigma^2)/\zeta\overset{\bD}{\to}N(0,1)$.
By rate condition $nV(\cP_\lambda f^{(0)}_*)=o(1)$, we get
$V(\widehat{f}^{\textrm{cs}}_{n\lambda}-f^{(0)}_*)=W_3+o_\Pr(\zeta)$.
Therefore, for any $u\in\bbR$,
\[
\Pr\left(V(\widehat{f}^{\textrm{cs}}_{n\lambda}-f^{(0)}_*)\le\sigma^2+u\zeta\right)\to\Phi(u).
\]
Proof is complete.
\end{proof}

\bibliographystyle{apalike}
\bibliography{ref}
\end{document}